\documentclass[apsprl,longbibliography,showpacs,floatfix,superscriptaddress,twocolumn]{revtex4-1}

\usepackage{graphicx,color}
\usepackage{amsfonts}
\usepackage[figuresright]{rotating}  
\usepackage{amssymb}
\usepackage{bm}
\usepackage{amsmath}
\usepackage{mathtools}
\usepackage{psfrag}
\usepackage{multirow}
\usepackage{tabularx}
\usepackage{textcomp}
\usepackage{units}
\usepackage{lipsum}

\usepackage{physics}
\usepackage{soul}
\usepackage{titlesec}
\usepackage{times}
\usepackage[colorlinks=true,citecolor=blue,linkcolor=magenta,hypertexnames=false]{hyperref}
\usepackage{enumitem}

\usepackage{subfigure}

\DeclareMathAlphabet\mathbfcal{OMS}{cmsy}{b}{n}

\titlespacing*{\section}
{0pt}{1ex plus .25ex}{1ex plus 1ex}
\titlespacing*{\subsection}
{0pt}{1ex plus .25ex}{1ex plus 1ex}
\titlespacing*{\subsubsection}
{0pt}{1ex plus .25ex}{1ex plus 1ex}

\titleformat{\section}
 {\fontsize{10}{15}\bfseries }
  {\thesection}
  {1em}
  {}

\titleformat{\subsection}
  {\bfseries }
  {\thesubsection}
  {2em}
  {}

\titleformat{\subsubsection}
  {\itshape}
  {\thesubsubsection}
  {2em}
  {}

\def\beq{\begin{eqnarray}}
\def\eeq{\end{eqnarray}}

\renewcommand{\v}[1]{\ensuremath{\mathbf{#1}}} % for vectors
 
\let\baraccent=\= % rename builtin command \= to \baraccent
\renewcommand{\=}[1]{\stackrel{#1}{=}} % for putting numbers above =
\newcommand{\mc}[1]{\mathcal{ #1}} % mathcal
\makeatletter

\titleclass{\subsubsubsection}{straight}[\subsection]

\newcounter{subsubsubsection}[subsubsection]
\renewcommand\thesubsubsubsection{\thesubsubsection.\arabic{subsubsubsection}}
\titleformat{\subsubsubsection}
   {\normalfont\normalsize\itshape \centering}{\thesubsubsubsection}{1em}{}
\titlespacing*{\subsubsubsection}
{0pt}{1ex plus .3ex}{1ex plus .3ex}
\makeatletter
\renewcommand\paragraph{\@startsection{paragraph}{5}{\z@}%
  {3.25ex \@plus1ex \@minus.2ex}%
  {-1em}%
  {\normalfont\normalsize}}
\renewcommand\subparagraph{\@startsection{subparagraph}{6}{\parindent}%
  {3.25ex \@plus1ex \@minus .2ex}%
  {-1em}%
  {\normalfont\normalsize}}
\def\toclevel@subsubsubsection{4}
\def\toclevel@paragraph{5}
\def\toclevel@paragraph{6}
\def\l@subsubsubsection{\@dottedtocline{4}{7em}{4em}}
\def\l@paragraph{\@dottedtocline{5}{10em}{5em}}
\def\l@subparagraph{\@dottedtocline{6}{14em}{6em}}
\makeatother

\begin{document}

\title{Time-reversal invariant topological skyrmion phases}
\author{R. Flores-Calderon}
%\affiliation{Max Planck Institute for Chemical Physics of Solids, Nöthnitzer Strasse 40, 01187 Dresden, Germany}
\affiliation{Max Planck Institute for the Physics of Complex Systems, Nöthnitzer Strasse 38, 01187 Dresden, Germany}
\affiliation{Max Planck Institute for Chemical Physics of Solids, Nöthnitzer Strasse 40, 01187 Dresden, Germany}

\author{Ashley M. Cook}
\affiliation{Max Planck Institute for the Physics of Complex Systems, Nöthnitzer Strasse 38, 01187 Dresden, Germany}
\affiliation{Max Planck Institute for Chemical Physics of Solids, Nöthnitzer Strasse 40, 01187 Dresden, Germany}

\begin{abstract}

Topological phases realized in time-reversal invariant (TRI) systems are foundational to experimental study of the broader canon of topological condensed matter as they do not require exotic magnetic orders for realization. We therefore introduce topological skyrmion phases of matter realized in TRI systems as a foundational step towards experimental realization of topological skyrmion phases. A novel bulk-boundary correspondence hidden from the ten-fold way classification scheme is revealed by the presence of a non-trivial value of a $\mathbb{Z}_2$ spin skyrmion invariant. This quantized topological invariant gives a finer description of the topology in 2D TRI systems as it indicates the presence or absence of robust helical edge states for open boundary conditions, in cases where the $\mathbb{Z}_2$ invariant computed with projectors onto occupied states takes a trivial value. Physically, we show this hidden bulk-boundary correspondence derives from additional spin-momentum-locking of the helical edge states associated with the topological skyrmion phase. ARPES techniques and transport measurements can detect these signatures of topological spin-momentum-locking and helical gapless modes. Our work therefore lays the foundation for experimental study of these phases of matter.

 \end{abstract}
\maketitle
\section{Introduction}

Topological phases of matter have challenged our understanding since the discovery of the integer and fractional quantum Hall effects \cite{QHE-VonKlitzing,FQH-exper,FQH-Laughlin} and the prediction of the paradigmatic Chern insulator \cite{chern-discovery,Haldane-model} culminating in the experimental realization of the quantum spin Hall insulator (QSHI) \cite{kane2005quantum,QSHI-HgTe-Theory,QSHI-HgTe-Exp,kane2005z2,graph-TI,edgestates-QSHI,QSHI-ham,Min2006IntrinsicAR,qian_2014} followed by the 3D TI \cite{Fu-Kane-Mele,3DTI-exp2,noh2018topological,peterson2018quantized,imhof2018topolectrical,serra2018observation, 3DTI-ham}. Due to the natural TR symmetry of most  experimental setups many new materials with TR-invariant topology have been identified such as in  TI ultra-thin films \cite{TI-thin-film-conduc,TI-crossover,TI-thin-phase-trans}, Van der Waals heterostructures \cite{vanwaals-hetero,Nature-hetero-antiferr,TI-hetero,robust-2dhetero}, and transition-metal dichalcogenides in particular given their large spin-orbit coupling \cite{TMD-1,TMD-topo}. These exotic states of matter present distinct bulk and edge properties generically characterized by topological invariants \cite{Kitaev-K-theory-10fold, 10-fold-way, WeylArmitage, McGinleyLindbladian,McGinley1D,essin2009,BalentsWeylTI}. In the case of effectively non-interacting topological phases, it is well known \cite{Moorehomotopy2007,ten-fold-way,10-fold-way} that a non-trivial topological invariant characterizes the topological phase in the bulk and implies the presence or absence of topologically robust metallic edge states which persist up to closing of the minimum direct bulk energy gap when the symmetries protecting the topological phase are respected \cite{HOTI,kane2005quantum,kane2005z2,noh2018topological,peterson2018quantized,imhof2018topolectrical,serra2018observation}. These topological phases correspond to topologically non-trivial mappings from the full Brillouin zone (BZ) to the space of projectors onto occupied states \cite{Moorehomotopy2007,ten-fold-way,10-fold-way,Kitaev-K-theory-10fold,Kitaev_2001}, with the number of topologically-distinct sectors defined by homotopy groups. An equivalent classification scheme is also derived by examining the nonlinear sigma models characterizing the topologically-protected states localized at the boundary, with explicit calculation of homotopy groups facilitated by K-theory~\cite{Kitaev-K-theory-10fold,Kitaev_2001}. 

The set of mappings from the full Brillouin zone to the space of an observable $\mathcal{O}$ generally exhibits topological sectors if the mappings correspond to a non-trivial homotopy group, however~\cite{First-skyrmion, Ashley-Shuwei}. The expectation value of $\mathcal{O}$ corresponding to such a non-trivial homotopy group can wind over the Brillouin zone to yield quantized, non-trivial topological charge for mappings in topological sectors, this topological charge is, in general, distinct from the topological charge due to winding of the projectors onto occupied states. A physically relevant degree of freedom that has been studied in previous works is spin \cite{First-skyrmion,Ashley-Shuwei,Prodan09-robustspin}, though only recent work~\cite{First-skyrmion,Ashley-Shuwei} considers cases where the spin topological invariant is not locked in value to the projector topological invariant~\cite{Prodan09-robustspin}. These topological phases associated with the spin degree of freedom specifically are known as topological skyrmion phases of matter~\cite{First-skyrmion}, which are now understood to be lattice counterparts and first evidence of quantized transport of magnetic skyrmions, a quantum skyrmion Hall effect~\cite{QSkHE}. Notably, the quantization of topological charge computed as the winding of the ground state spin expectation value over the full BZ is guaranteed whenever the minimum magnitude of the ground state spin expectation value is finite, even when spin is not a conserved quantity \cite{First-skyrmion,Ashley-Shuwei}. 
\section{Models}
In the following, we consider tight-binding Hamiltonians which preserve time-reversal symmetry (TRS) and realize skyrmions in spin textures over the Brillouin zone. To construct such Hamiltonians, we combine a TRS-breaking Hamiltonian, which realizes a topological skyrmion phase, with its time-reversed partner, and couple these two Hamiltonians with additional spin-orbit coupling (SOC) terms. This approach takes inspiration from construction of some Hamiltonians describing quantum spin Hall insulator phases, by pairing a Chern insulator with its time-reversed partner, and then adding non-negligible SOC terms \cite{kane2005quantum,QSHI-HgTe-Theory,kane2005z2}. A similar construction was used to realize first examples of helical topological skyrmion phases~\cite{First-skyrmion} characterized by non-trivial skyrmion numbers computed for mirror subsectors, equal in magnitude and opposite in sign. The present work goes beyond this past study in that block-diagonalization is not required, and time-reversal symmetry is present. While topological skyrmion phases of matter are realized by three-band Bloch Hamiltonians~\cite{QSkHE}, the simplest cases of topological skyrmion phases are, arguably, in four band models with generalized particle-hole symmetry~\cite{First-skyrmion,Ashley-Shuwei}, we consider an eight band model. The additional degree of freedom has to be given by the spin of the electron since the four band models possess no TR symmetry. We therefore reinterpret the skyrmion for each spin sector as forming in the texture of a pseudo-spin degree of freedom over the Brillouin zone. This is consistent with past work on topological skyrmion phases: ultimately, any pseudo-spin with the appropriate set of deformations to yield a non-trivial homotopy group has the potential to realize a topological skyrmion phase~\cite{QSkHE}. 

Explicitly, we consider the following Hamiltonian for the TRI topological skyrmion phase:

\begin{align}
    H_1(\textbf{k})=&s_0(\sin{k_y}\tau_0 \sigma_y+\epsilon(k_x,k_y)\tau_z\sigma_z)+\sin{k_x}s_z\tau_z \sigma_x \notag \\&+\Delta(k_x,k_y)+V_{\text{SOC}}+V_{\text{bulk}}\label{Skyrm-mod1},
\end{align}

where $s_\mu,\tau_\mu,\sigma_\mu,\ \mu=0,1,2,3$ are Pauli matrices acting on the spin, particle-hole and pseudo-spin degrees of freedom respectively. We have defined $\epsilon(\textbf{k})=2+M-t(\cos{k_x}+\cos{k_y})$ where $t,M$ are real numbers describing hopping and staggered potential amplitudes. We further restrict ourselves to spin-triplet pairing in the pseudo-spin sector $\Delta(\textbf{k})=\Delta_0(\sin{k_x}s_0\tau_y\sigma_0-\sin{k_y}s_z\tau_x\sigma_z)$, where $\Delta_0$ is the pairing strength, and spin orbit coupling  term $V_{\text{SOC}}=cs_x\tau_0\sigma_y$. 

Given the high symmetry of the Hamiltonians, there is some possibility of other symmetries protecting other topological phases besides those characterized by the $\mathbb{Z}_2$ projector invariant of the QSHI, or the skyrmion invariant. We therefore also include the term $V_{\text{bulk}}$, which may contain additional perturbations to break all symmetries save for particle-hole symmetry and time-reversal symmetry as discussed below. 
\begin{figure}[ht]
    \includegraphics[width=\columnwidth]{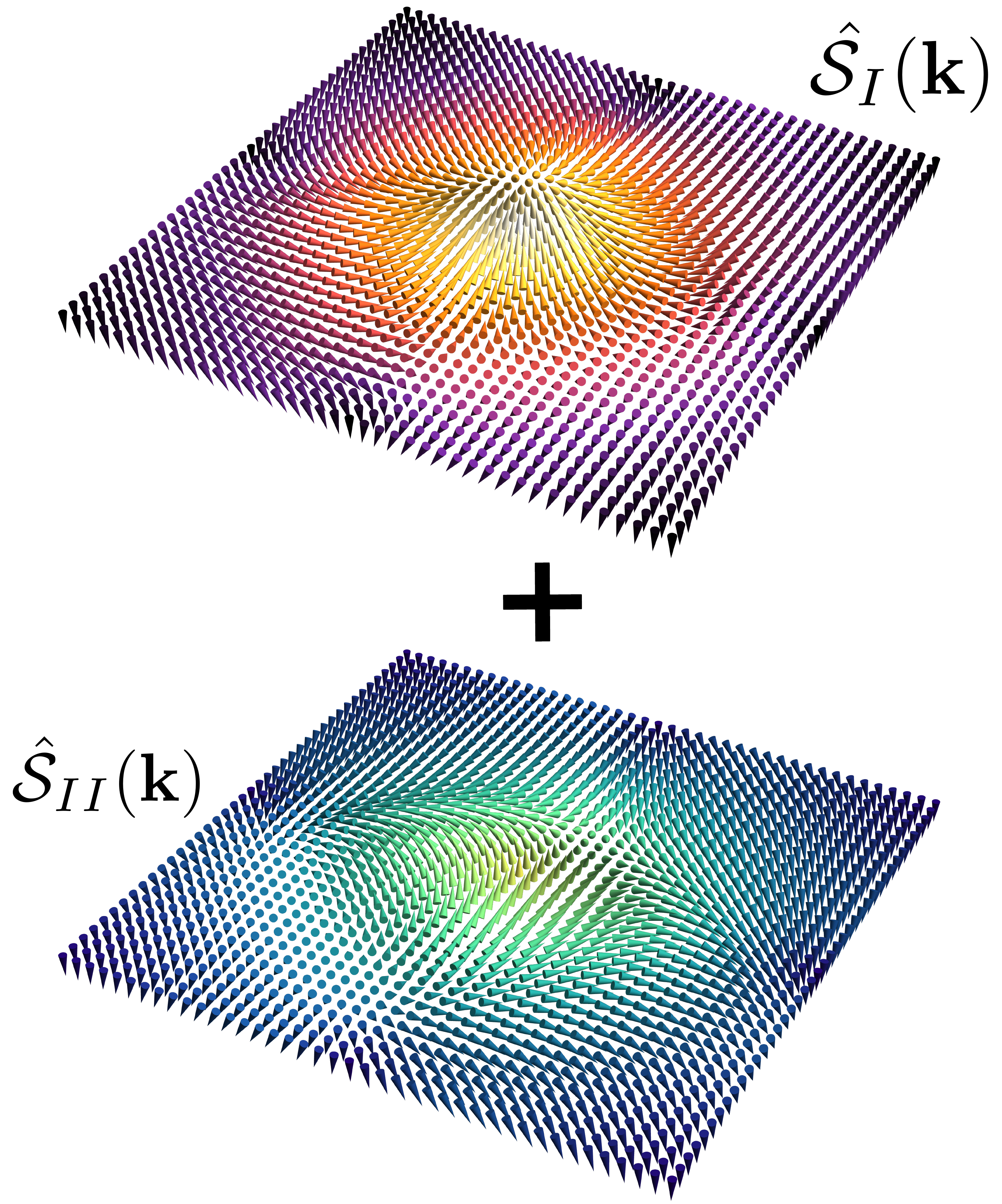}
  \vspace{-0.3cm}
  \caption{Pseudo-spin textures of the TR skyrmion phase analyzed in the main text. The arrows represent the pseudo-spin expectation value $\expval{S^I_\mu}_{\mathbf{k}}$ for one TR sector denoted by $I$, this skyrmion texture has a TR partner (lower skyrmion texture), which is given by the pseudo-spin expectation value $\expval{S^{II}_\mu}_{\mathbf{k}}$ of the $II$ TR sector. The color is proportional to the polar angle of the texture. The skyrmion numbers correspond to $\mathcal{Q}_{I}=+ 1,\ \mathcal{Q}_{II}=-1$ for the model of \eqref{Skyrm-mod1} with $M=-1$, $\Delta_0=0.5$,$\lambda=0.3$,$c=0.5$.  }
  \label{TR-Skyrm}
\end{figure}
For negligible SOC, this Hamiltonian consists of two topological skyrmion phase Hamiltonians, which possess total Chern number $\mathcal{C}_{tot}$ of $+2$ or $-2$ and skyrmion number $\mathcal{Q}$ of $-1$ or $+1$ for half-filling, respectively, depending on the spin sector. For non-negligible SOC, this corresponds to a trivial value for the $\mathbb{Z}_2$ projector topological invariant of the quantum spin Hall insulator phase, and non-trivial value for the $\mathbb{Z}_2$ skyrmion topological invariant.

To explore the consequences of the non-trivial skyrmion topology, we additionally consider a counterpart TRS Hamiltonian constructed from TRB topological skyrmion phase Hamiltonians with even total Chern number and even skyrmion number:

\begin{align}
    H_2(\textbf{k})=&s_0(t\cos{k_x}\tau_z\sigma_x+\cos{(k_x+k_y)}\tau_z\sigma_z)\notag \\&+t\cos{k_y}s_z \tau_0 \sigma_y+\Delta(k_x,k_y)+V_{\text{SOC}}\label{Skyrm-mod2},
\end{align}

As these two Hamiltonians each possess PHS, they can be written in the Nambu basis, to describe superconductors at mean-field level using Bogoliubov de Gennes formalism \cite{Bogoljubov1958,Valatin1958}. Eq.~\ref{Skyrm-mod1} and Eq.~\ref{Skyrm-mod2} are invariant under the particle-hole operators $\mathcal{C}_1=s_z\tau_x\sigma_z K$ and $\mathcal{C}_2=\tau_x K$, respectively, which each square to one. By construction, each model is also invariant under the spinful time-reversal operation $\mathcal{T}=-is_y K$.\\

%We find that these models possess a generalized particle-hole symmetry as first discussed in \cite{Ashley,Hopf} given by $\mathcal{C}^{'} =-i\tau_y K$ where $K$ represents complex conjugation. 

\section{Topological characterization}

To characterize the topology of these models, we employ a generalization of the skyrmion invariant to TRI-systems. The use of the total skyrmion number to characterize the topology fails, since, in analogy to the Chern number, it is always zero when spinful TRS is present, as proved in appendix S1. In the case of the quintessential 2D TR-invariant systems classified by the ten-fold way, one can construct a non-zero topological invariant in multiple equivalent formalisms. One such path involves the usage of Kramer's theorem to express the Hamiltonian in terms of the helicity basis or Kramers pair basis, where the Hamiltonian is block diagonal. This can be done in the case of negligible SOC.  Then, for each Hamiltonian one can calculate the Chern number and define a parity index $\nu=(C-C_{TR})/2\  \text{mod}(2)=(-1)^C$. The parity index is equivalent to the celebrated $\mathbb{Z}_2$ invariant of Kane-Mele \cite{kane2005quantum,kane2005z2}.\\

In analogy to this parity invariant, we may compute a skyrmion number for each spin sector, characterizing a topological pseudo-spin texture over the Brillouin zone. The explicit form of the spin operators is given in appendix S2 and a schematic visualization of this mapping is shown in Fig.~\ref{TR-Skyrm}. Once these operators are defined, the topological skyrmion invariant can be calculated for each spin sector so as to obtain two $\mathcal{Q}_{I},\mathcal{Q}_{II}\in \mathbb{Z}$ integers. As detailed in appendix S2, these quantities are opposite in sign and equal in magnitude so that it is natural to define a \textit{TR-skyrmion invariant} or \textit{skyrmion parity} given by $\nu_{\mathcal{Q}}=(\mathcal{Q}_I-\mathcal{Q}_{II})/2\  \text{mod}(2)$. This invariant, which now fully takes into account the TR-invariant nature of the system, will be shown to be linked to a novel bulk-boundary correspondence when open boundary conditions in $x$ or $y$ are applied. It will be further shown numerically that the helical edge states present when the TR-skyrmion invariant is non-trivial are robust to disorder and local perturbations. Furthermore, the remarkable behaviour displayed by this phase doesn't rely on crystalline point group symmetries, as will be shown by explicitly breaking all symmetries except for time reversal and particle-hole.

\section{Results}

Starting with the simplest case of negligible spin-orbit coupling, we consider two uncoupled Hamiltonians labeled by spin sector, which each possess four bands, with a well-defined Chern number and skyrmion number as described in Liu~\emph{et al.}~\cite{Ashley-Shuwei}. For $H_1(\textbf{k})$ the total Chern number of each sector takes only two non trivial values $C_I=\pm2$ and $C_{II}=-C_I$ because of TRS. As the topological invariant for a quantum spin Hall insulator with non-negligible SOC according to the ten-fold way classification scheme is $\nu=(C_I-C_{II})/2\  \text{mod}(2)$, the value of $\nu$ for our Hamiltonian is then always trivial as $\nu= 2\  \text{mod}(2)= 0 \ \text{mod} \ 2$.

In contrast, $\mathcal{Q}_I=\mp 1$ in the regions of the phase diagram where $C_I = \pm 2$. Therefore, the skyrmion number for the TRI system, $\nu_{\mathcal{Q}}$, evaluates to $\nu_{\mathcal{Q}}=(\mathcal{Q}_I-\mathcal{Q}_{II})/2\  \text{mod}(2)= 1\  \text{mod}(2)$ for these regions of the phase diagram. The system with non-negligible SOC is therefore in a topologically non-trivial phase according to the skyrmion invariant, even when the projector invariant indicates the system is topologically trivial.

To explore the consequences of the non-trivial skyrmion number paired with trivial projector topological invariant, we consider the case of trivial projector invariant and trivial skyrmion number by comparing results for the Hamiltonian given by Eq.~\ref{Skyrm-mod1} with those for the Hamiltonian given by Eq.~\ref{Skyrm-mod2}. In the case of Eq.~\ref{Skyrm-mod2}, $C_I=\pm4$ in the topologically non-trivial regions of the phase diagram, corresponding to $\nu=0$ when SOC is non-negligible, and $\mathcal{Q}_I=\mp 2$, yielding a trivial $\nu_{\mathcal{Q}}=0$ as well. Comparing these two cases, we observe additional structure that is captured by the skyrmion invariant, which is richer than the ten-fold way classification scheme.\\

The previous analysis is valid when no SOC term is present. For $\nu$ even, it is known that as long as $V_{\text{SOC}}$ doesn't break TRS and doesn't close the minimum direct bulk energy gap, then the invariant will always be trivial and the system is expected to be topologically trivial. We therefore explore the effects of non-negligible SOC in systems characterized by the $\nu$ invariant, but also the previously-unidentified skyrmion invariant $\nu_{\mc{Q}}$.

\begin{figure}[ht]
    \subfigure[]{\includegraphics[width=0.49\columnwidth]{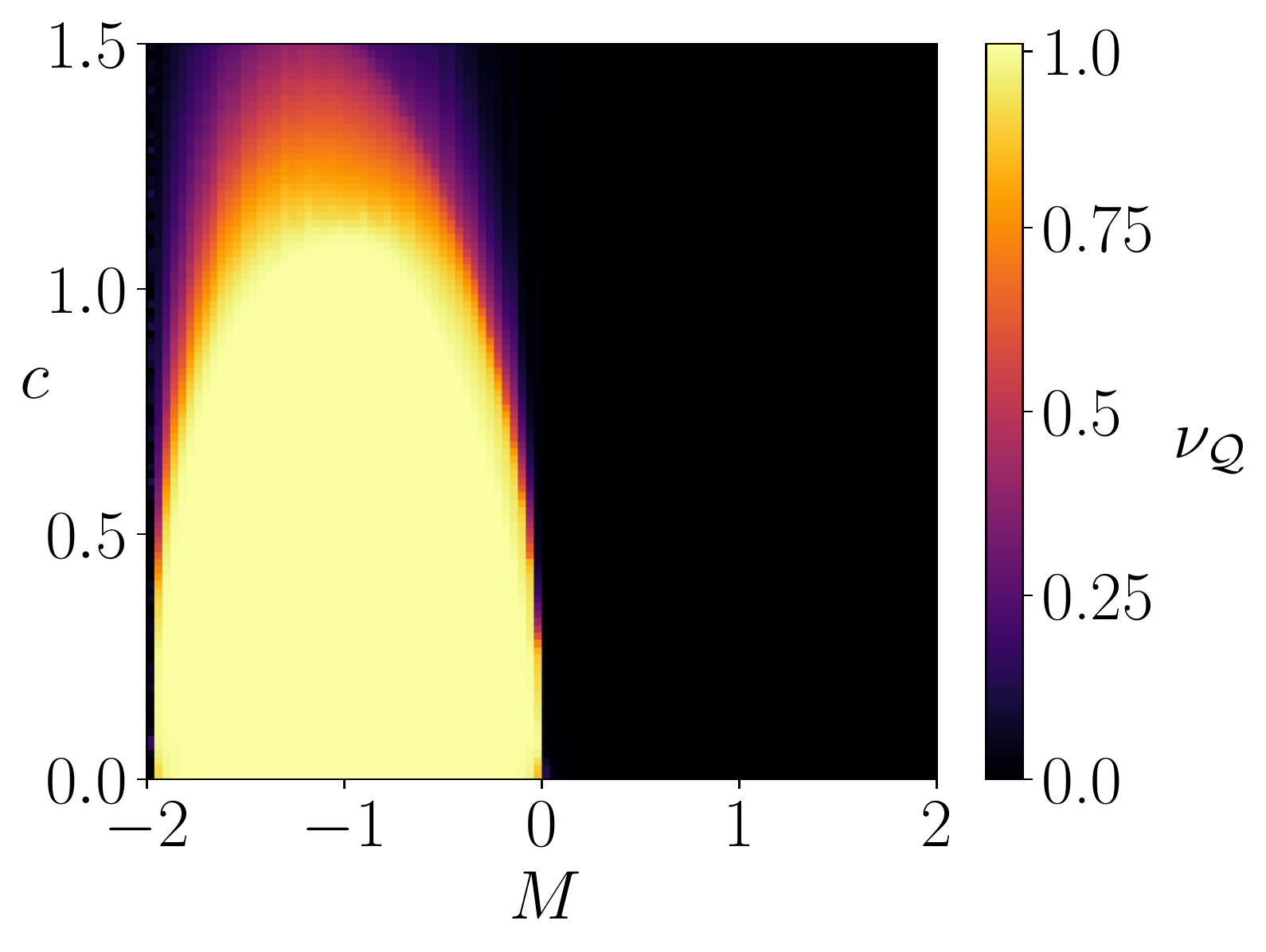}}
     \subfigure[]{\includegraphics[width=0.49\columnwidth]{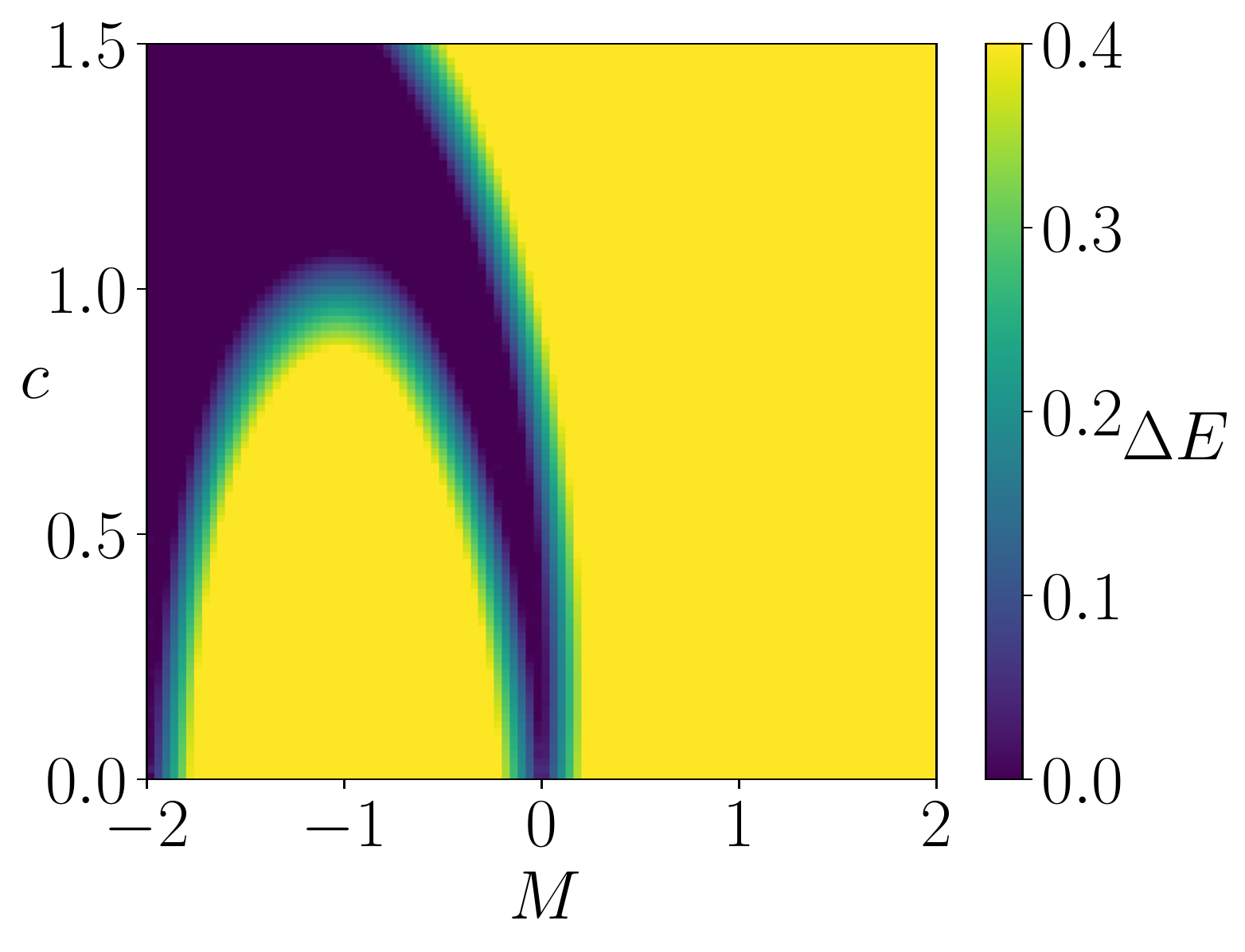}}
     \subfigure[]{\includegraphics[width=0.49\columnwidth]{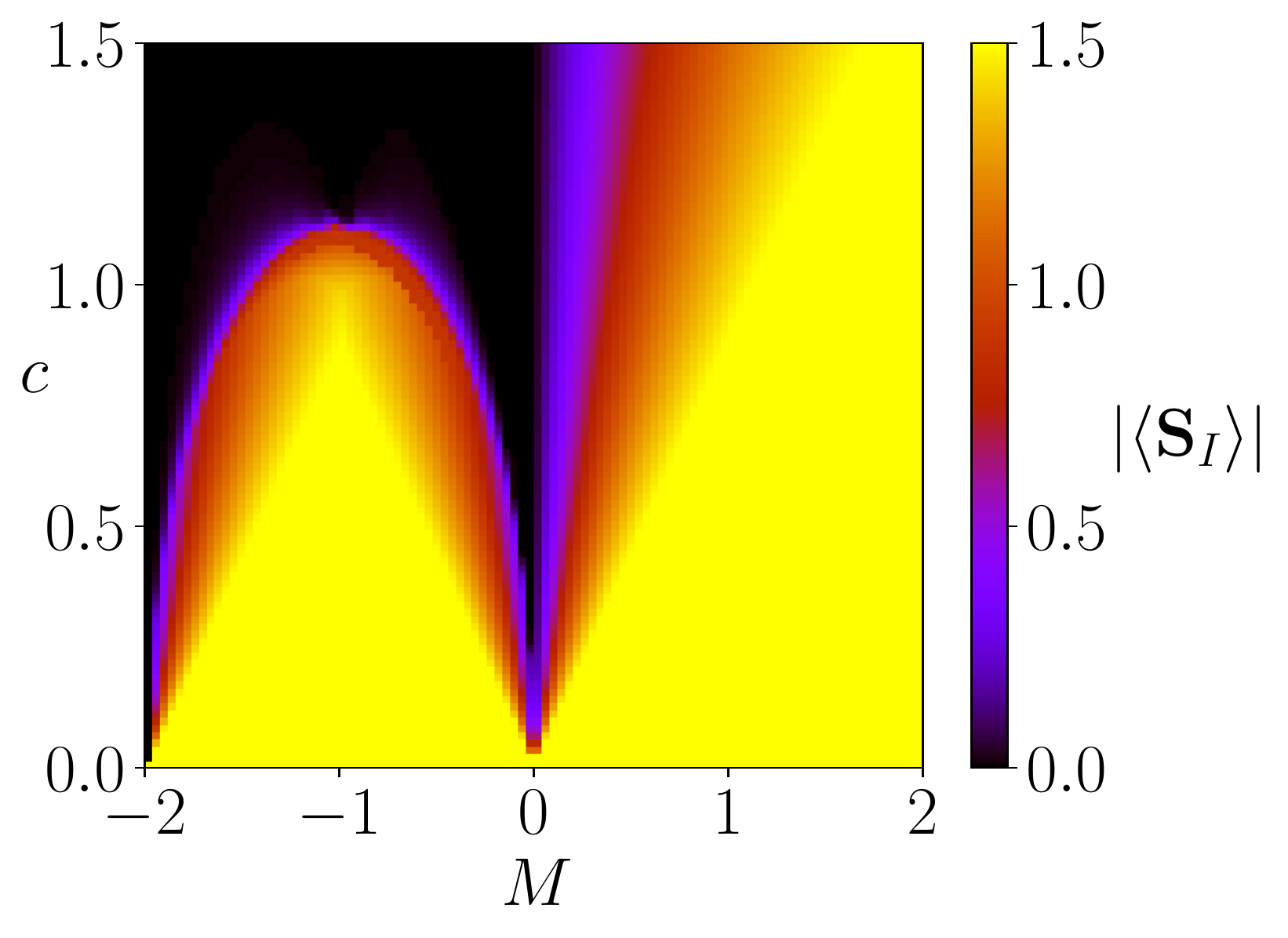}}
     \subfigure[]{\includegraphics[width=0.49\columnwidth,height=3.2cm]{     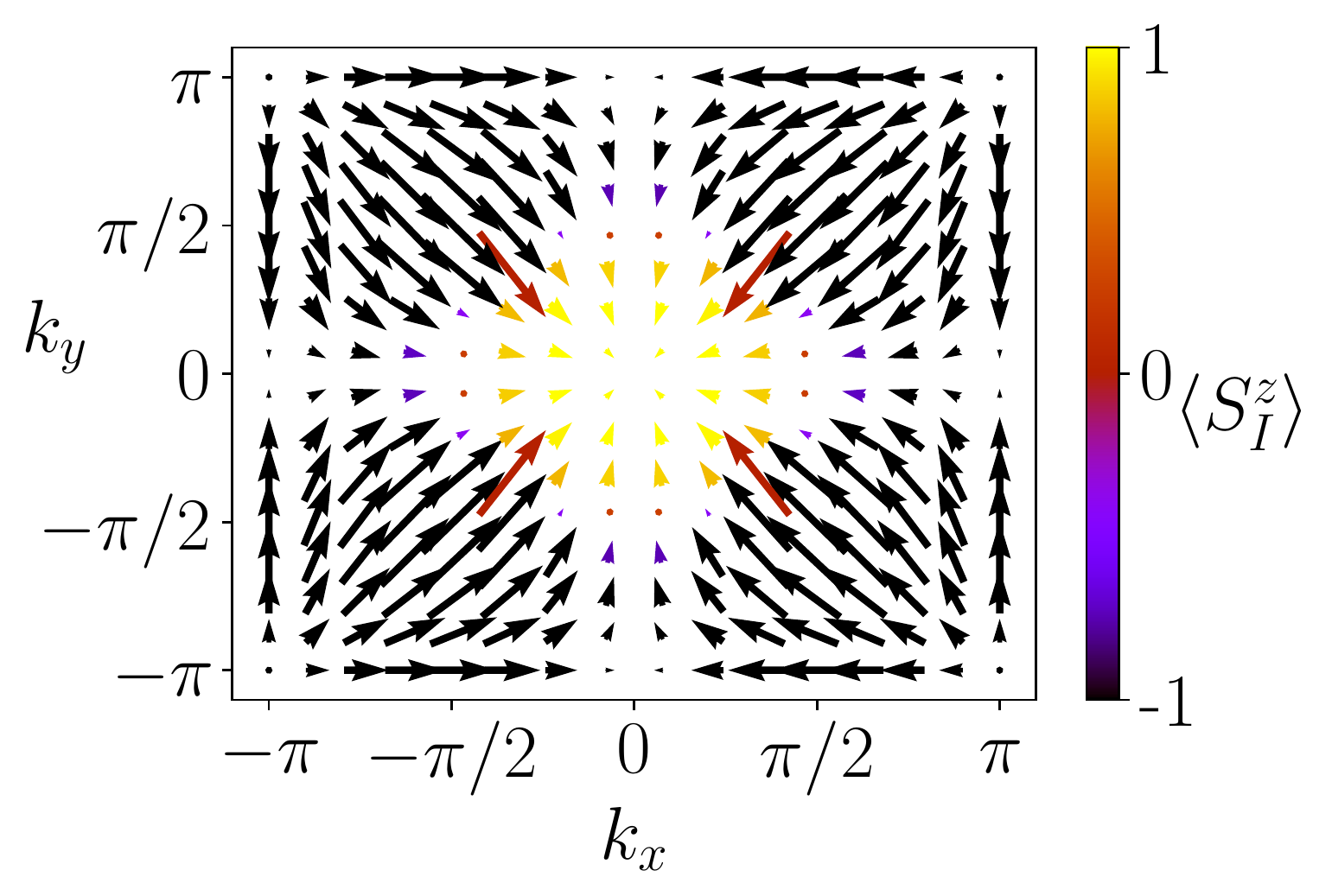}}

  \vspace{-0.3cm}
  \caption{a) Phase diagram of the TR-skyrmion invariant $\nu_\mathcal{Q}$ for a fix value of $\Delta_0=0.5$ and bulk perturbation $\lambda=0.3$ of model \eqref{Skyrm-mod1}.b) Minimum direct gap as a function of $c$ and $M$ for the same parameter values c) Minimum pseudo-spin magnitude for the first TR or spin up sector. d) skyrmion texture for vanishing magnitude $c=1.5$,$M=-1$,for the first TR sector, the arrows represent the vector $(\langle S^x_I \rangle,\langle S^y_I \rangle)$ while the color represents $\langle S^z_I \rangle$ }
  \label{skyrm-phase}
\end{figure}

Having defined this TR-skyrmion invariant $\nu_{\mc{Q}}$, we compute phase diagrams characterizing the skyrmion topology for Hamiltonian Eq.~\ref{Skyrm-mod1}, which are shown in Fig.~\ref{skyrm-phase}. $\nu_{\mc{Q}}$ is shown in Fig.~\ref{skyrm-phase} a) for  $\Delta_0=0.5$, as a function of SOC constant $c$ and mass parameter $M$. To understand the stability of the TRI topological skyrmion phase and corresponding quantization of $\nu_{\mc{Q}}$, we also show the minimum direct bulk energy gap $\Delta E$ in Fig.~\ref{skyrm-phase} b) and the minimum spin magnitude 
$|\langle S_{I} \rangle |$ in Fig.\ref{skyrm-phase} c), respectively, each as a function of $c$ and $M$. For this parameter set, two regimes are distinguished in the case of $H_1(\textbf{k})$: the trivial regime of zero skyrmion number, and the region of $\mathcal{Q}_I=\pm 1,\mathcal{Q}_{II}=\mp 1$. Quantization of $\nu_{\mc{Q}}$ survives for finite $c$ and is non-trivial for the regime $-2<M<0$ when each of $\Delta E$ and $|\langle S_{I} \rangle |$ are finite. For sufficiently large $c$, $|\langle S_{I} \rangle |$ goes to zero, and $\nu_{\mc{Q}}$ is finite but unquantized, smoothly approaching zero with further increase of $c$. The pseudospin texture corresponding to one of these unquantized values of $\nu_{\mc{Q}}$ is shown in Fig.~\ref{skyrm-phase} d). This situation is analogous to the loss of quantization of the Hall conductivity in a Chern insulator when the Fermi level intersects bands, rather than bands being completely filled or empty. 

For this model, $\nu_{\mc{Q}}$ only changes when $\Delta E = 0$ corresponding to a type-I topological phase transition.  This corresponds to symmetry-protection of two sets of spin operators, each with $4\times 4$ matrix representation, as opposed to a lower-symmetry model with a single set of spin operators with $8 \times 8$ matrix representations: we have effectively two four-band toy models for topological skyrmion phases still despite finite SOC terms, and it is previously known that four-band models for topological skyrmion phases with generalized particle-hole symmetry $\mc{C}'$ lack sufficient degrees of freedom to realize the type-II topological phase transition~\cite{Ashley-Shuwei}, although three-band models without $\mc{C}'$ symmetry realize type-II topological phase transitions quite generically~\cite{QSkHE}.

\begin{figure}[ht]
    \subfigure[]{\includegraphics[width=0.49\columnwidth,height=3.5cm]{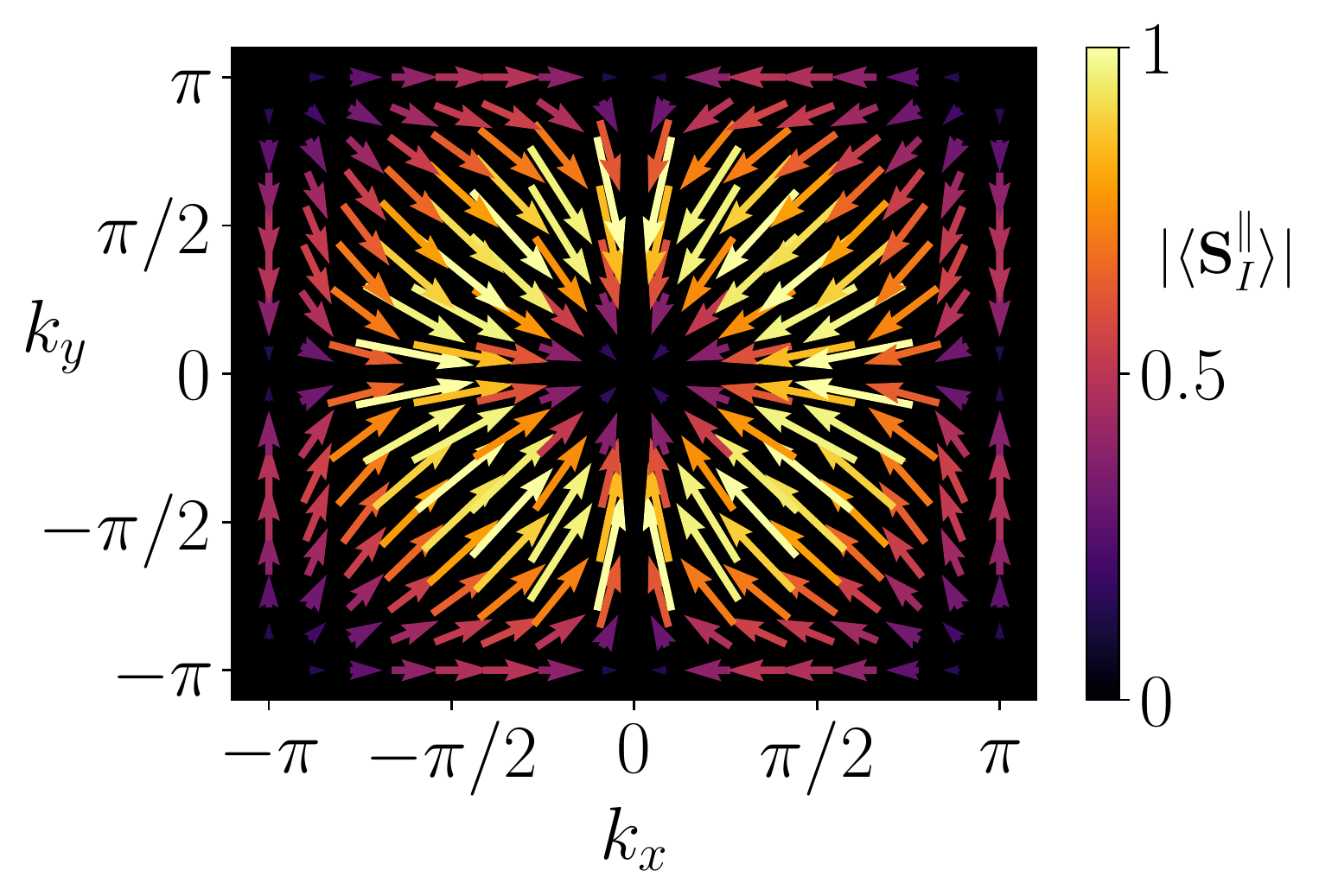}}
    \subfigure[]{\includegraphics[width=0.49\columnwidth,height=3.5cm]{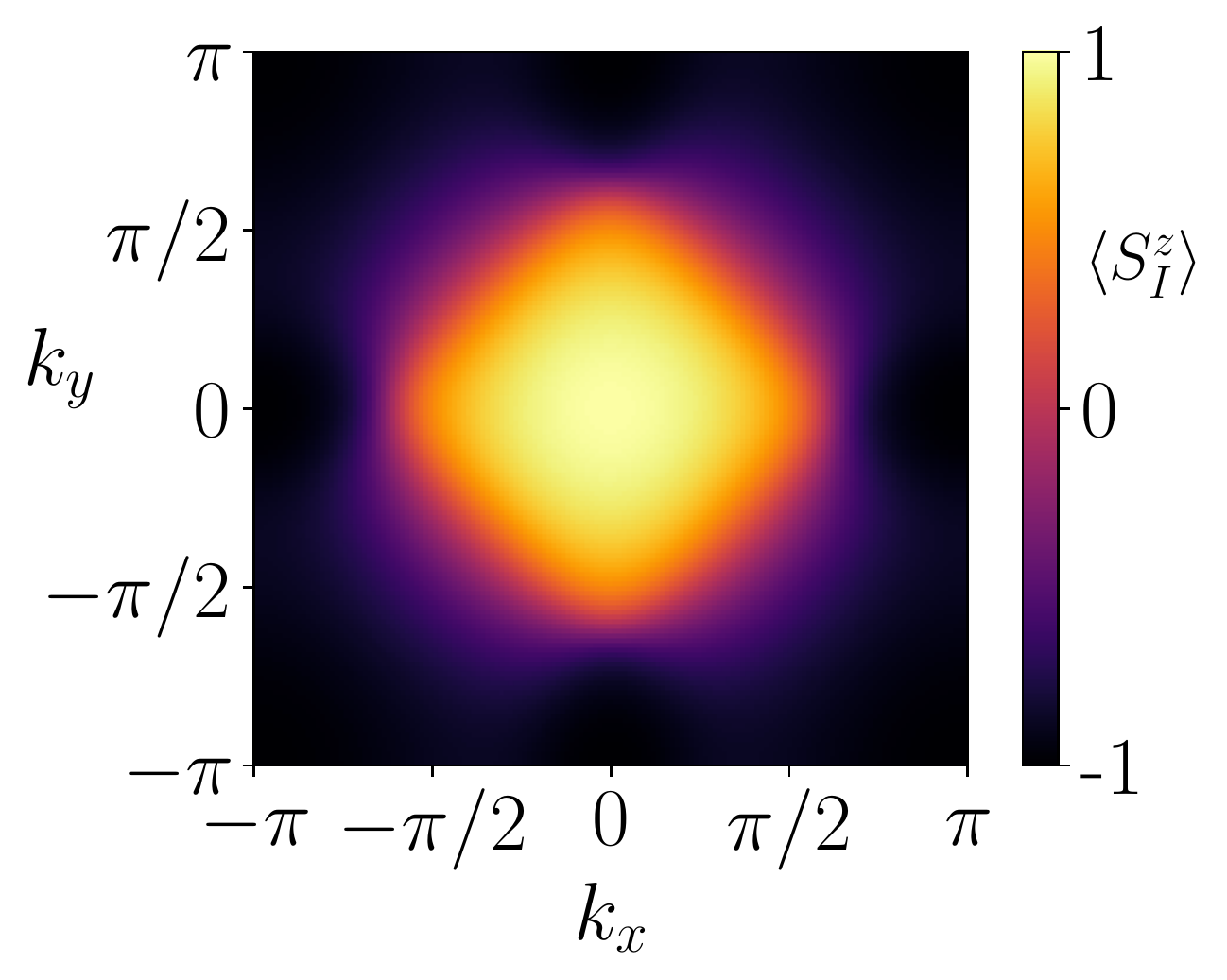}} 
    \subfigure[]{\includegraphics[width=0.49\columnwidth,height=3.5cm]{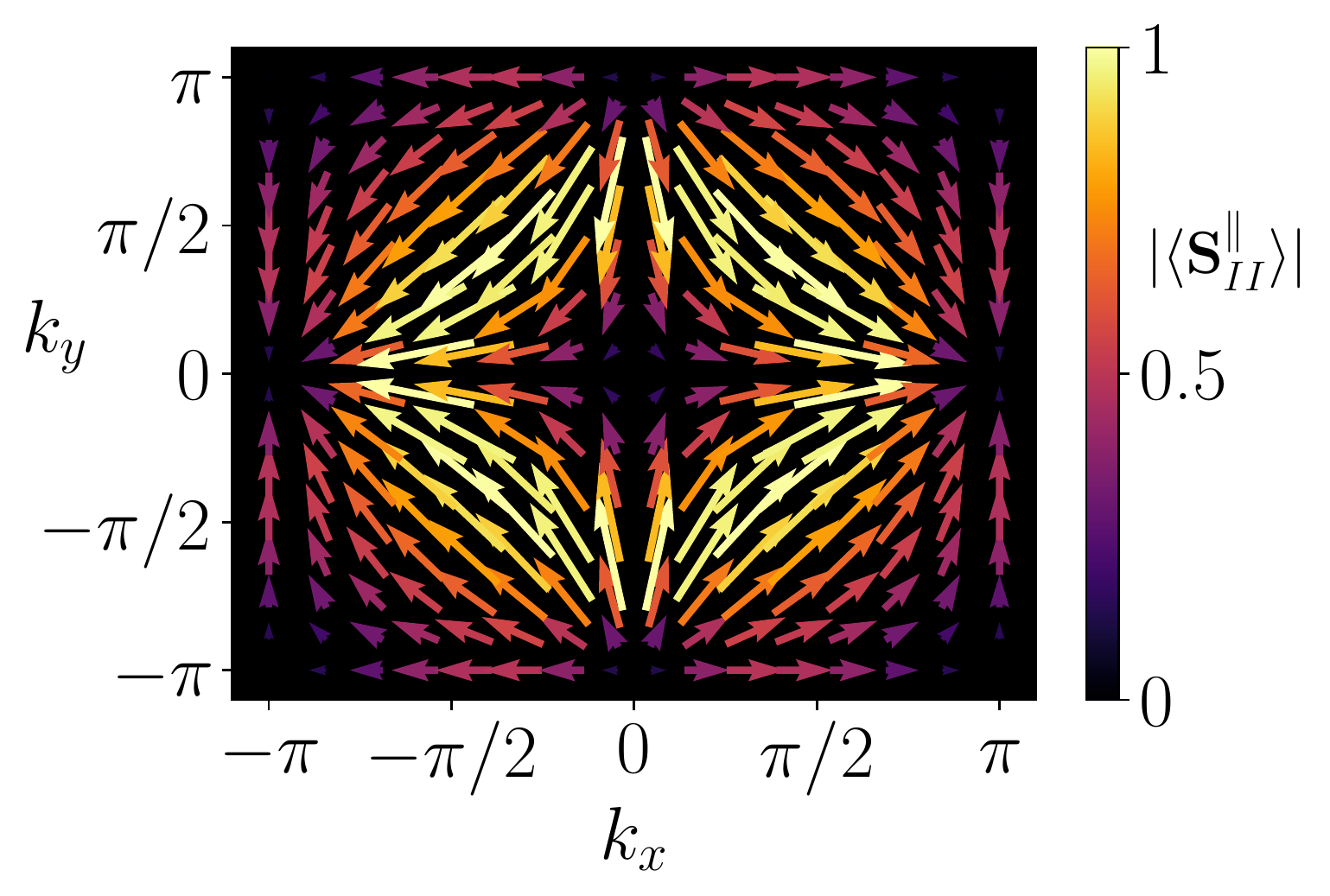}}
    \subfigure[]{\includegraphics[width=0.49\columnwidth,height=3.5cm]{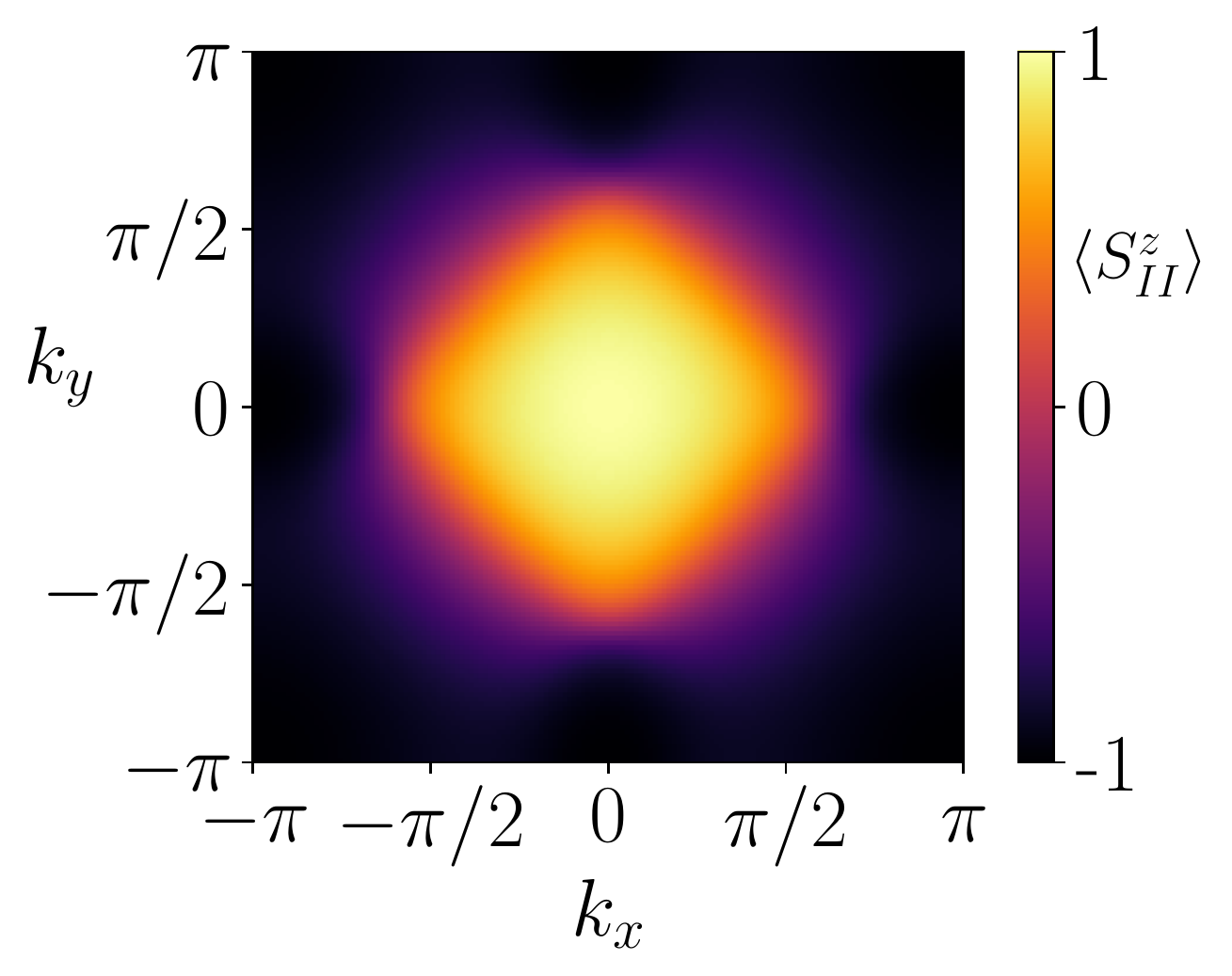}} 
  \vspace{-0.3cm}
  \caption{ Normalized ground state pseudo-spin expectation value plots for the model of \eqref{Skyrm-mod1} with $M=-1.3$, $\Delta_0=0.5$,$\lambda=0.4$,$c=0.8$. a) and c) show the $x-y$ pseudo-spin texture and normalized  $\lvert \langle (\mathcal{S}^{I}_x(\mathbf{k}),\mathcal{S}^{I}_y(\mathbf{k})) \rangle\rvert $,$\lvert \langle(\mathcal{S}^{II}_x(\mathbf{k}),\mathcal{S}^{II}_y(\mathbf{k})) \rangle\rvert $ respectively. Subfigures b) and d) show the normalized $z$ component of the pseudo-spin texture $ \langle \textbf{S}^z_{I} \rangle $ and $ \langle \textbf{S}^z_{II} \rangle $ respectively.}
  \label{Skyrm-textures}
\end{figure}

An example of the skyrmion texture with $\mathcal{Q}_I=+ 1$, with $M=-1.3$, $\Delta_0=0.5$,$c=0.8$ is shown in Fig.~\ref{Skyrm-textures} a),b). The time reverse partner with the antiskyrmion $\mathcal{Q}_{II}=- 1$ is shown in Fig.~\ref{Skyrm-textures} c), d) for the same parameter values. 

We now include the effect of a symmetry-breaking term additionally to the normal SOC given by:

\begin{align}
    V_{\text{bulk}}(\textbf{k})=\lambda(\sin(k_x+k_y)\cos(k_y)s_y\tau_x\sigma_0+\sin(k_x)s_x\tau_z\sigma_0)
\end{align}

This extra term in the Hamiltonian breaks all spurious crystalline symmetries of the original Hamiltonian in equation \eqref{Skyrm-mod1}. Included in those symmetries is inversion so that inversion is broken in the bulk as well as $\mathcal{C}'$ symmetry. The only remaining symmetries in the model are particle-hole $\mathcal{C}$ and time-reversal $\mathcal{T}$ symmetry as well as the combined chiral $\mathcal{C}\mathcal{T}$.

Fig.~\ref{Skyrm-textures} shows a winding within each of the two TR sectors (I and II) that remains even with non-negligible SOC. \\

\begin{figure}[ht]
    \subfigure[]{\includegraphics[width=0.49\columnwidth,height=3.5cm]{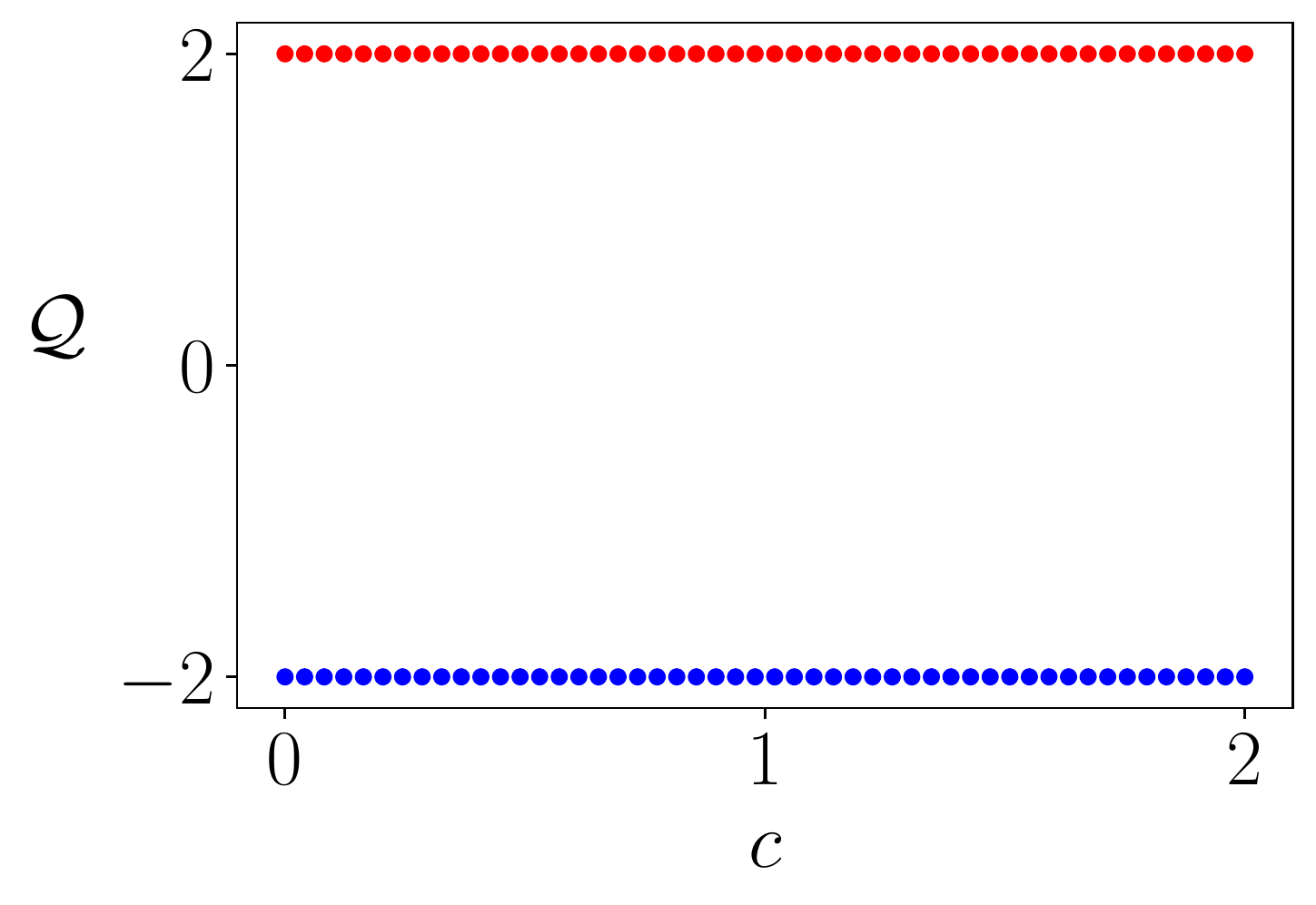}}
    \subfigure[]{\includegraphics[width=0.49\columnwidth,height=3.5cm]{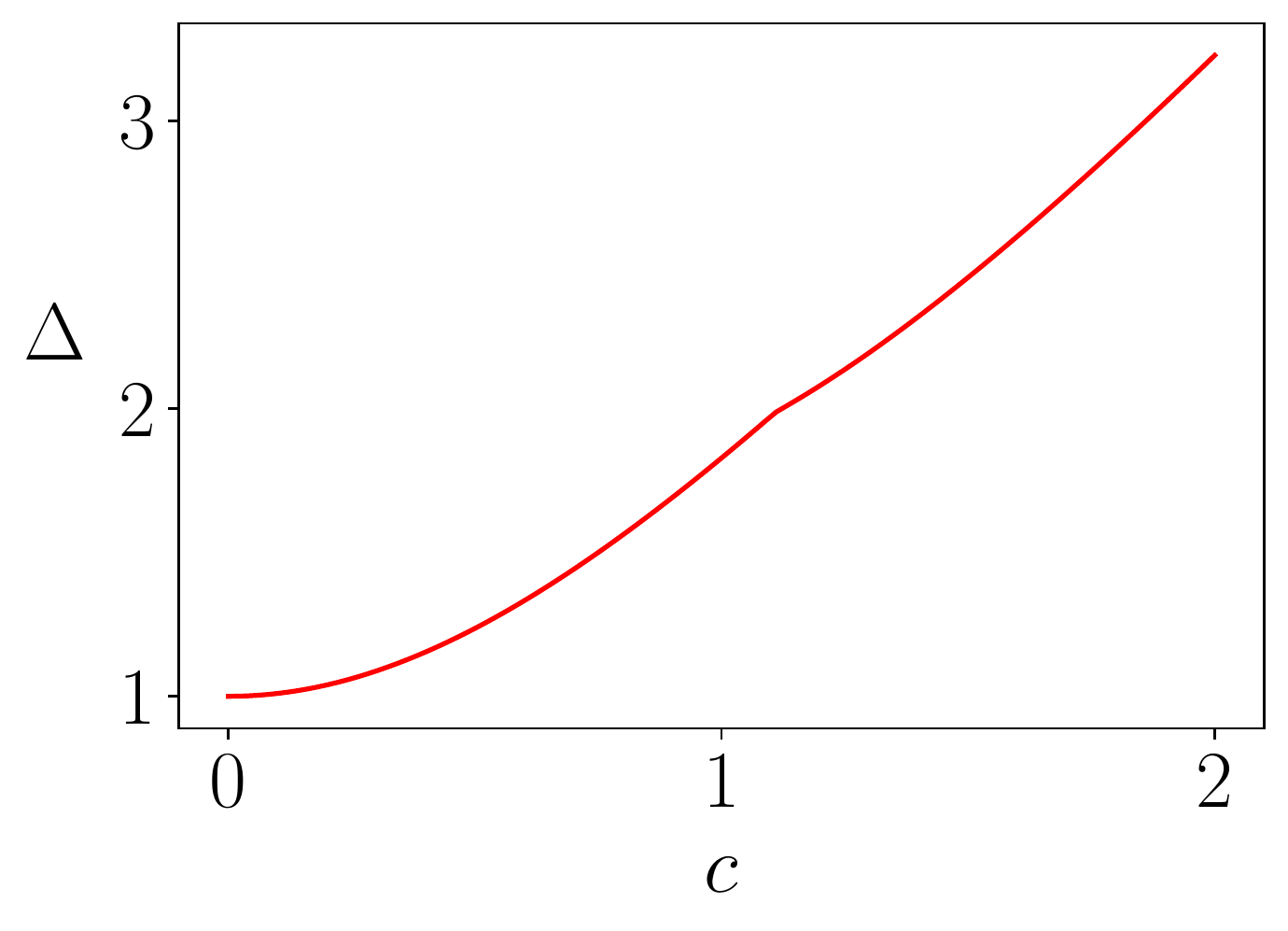}}
    \subfigure[]{\includegraphics[width=0.49\columnwidth,height=3.5cm]{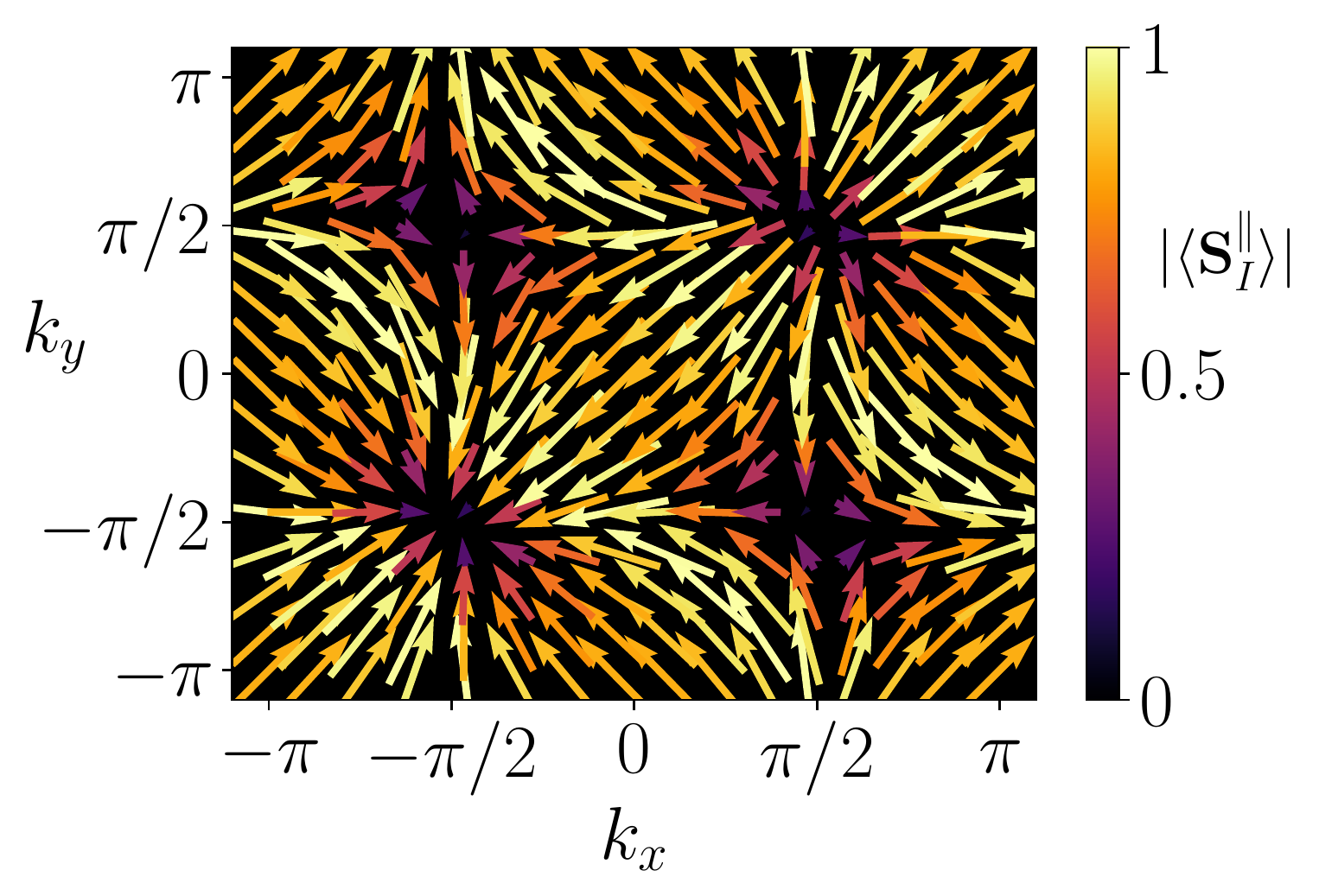}}
    \subfigure[]{\includegraphics[width=0.49\columnwidth,height=3.5cm]{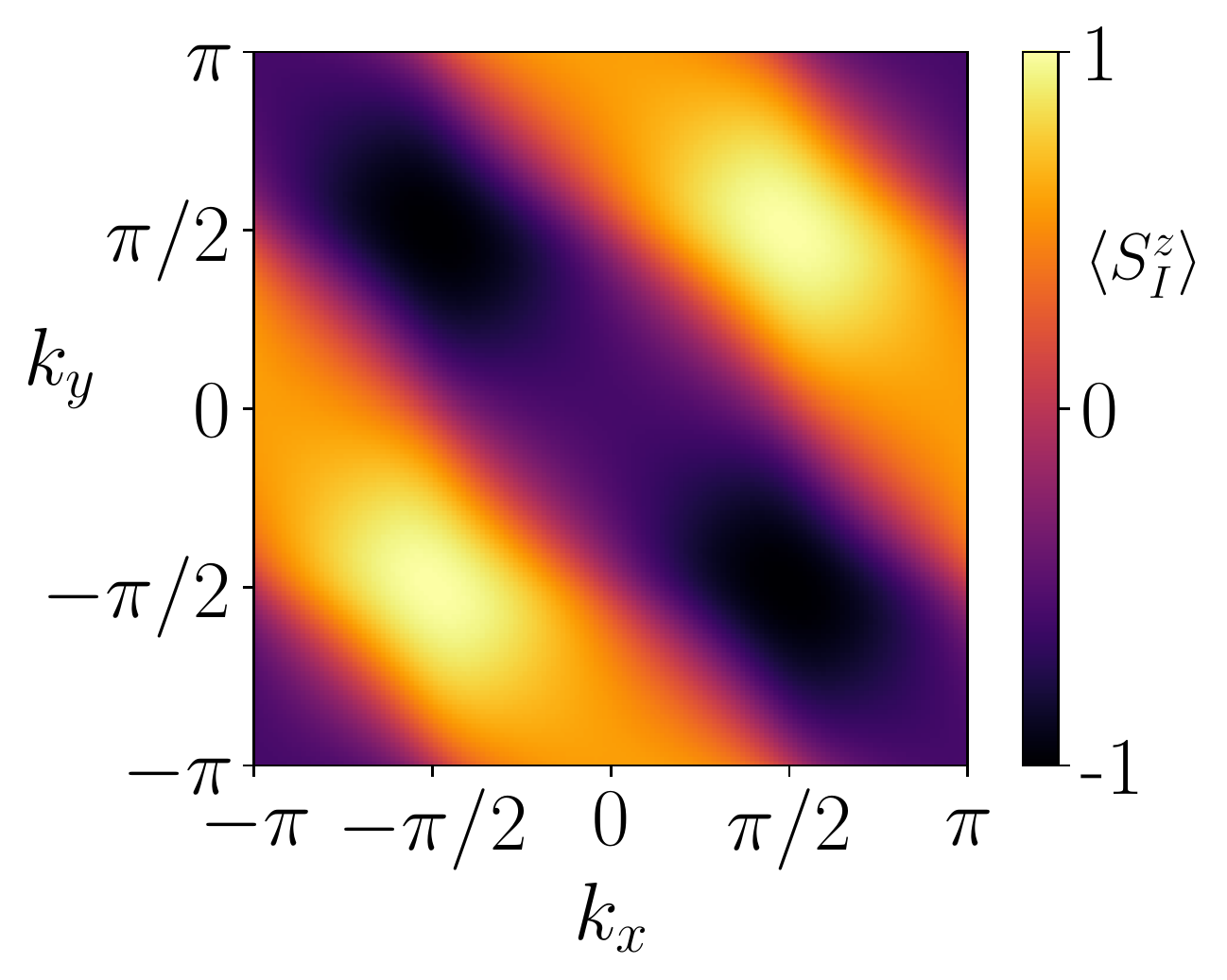}}
  \vspace{-0.3cm}
  \caption{a) Skyrmion numbers $\mathcal{Q}_I$ (red) and $\mathcal{Q}_{II}$ (blue) computed for $M=1,\Delta_0=0.5$ as a function of SOC $c$ for the model Hamiltonian of equation \eqref{Skyrm-mod2}. b) Minimum direct gap between bands enclosing zero Fermi energy. c), d) Normalized ground state pseudo-spin expectation value plots with $t=1$, $\Delta_0=0.5$,$\lambda=0.4$,$c=0.5$. Subfigure c) shows the $x-y$ pseudo-spin texture  $\lvert \langle (\mathcal{S}^{I}_x(\mathbf{k}),\mathcal{S}^{I}_y(\mathbf{k})) \rangle\rvert $, while Subfigures d) shows the normalized $z$ component of the pseudo-spin texture $ \langle \textbf{S}^z_{I} \rangle $}
  \label{Skyrm-numb-mod2}
\end{figure}

An important region of the phase diagram is the zero energy and zero pseudo-spin magnitude regime above approximately $c=1.2$. In this case because of the gapless spectrum, the skyrmion number destabilizes and $\mathcal{Q}_{I},\mathcal{Q}_{II}$ are not well defined just as the prototypical $\mathbb{Z}_2$ invariant. Remarkably, as shown in Fig \ref{skyrm-phase} d), the pseudo-spin still winds in such a manner that almost everywhere in the BZ it forms a skyrmion. The points of zero spin magnitude seem to not deform the global structure that give rise to the skyrmion texture, although the TR skyrmion invariant is destabilized in this case. It is worth noting that we can also do the same analysis for the second Hamiltonian $H_2(\textbf{k})$, in this case the pseudo-spin textures for each TR subsector result in a quantized skyrmion invariant of $\pm 2$ and thus not only a trivial projector invariant but also a trivial skyrmion parity. which is shown as a function of SOC in Fig. \ref{Skyrm-numb-mod2} along with an example of a skyrmion texture. It is worth mentioning that even if the skyrmion parity is trivial across the phase-diagram we still observe quantization of the skyrmion invariant as well as the non-trivial relation $Q_I=-Q_{II}$ holding for all parameter values.

\section{Bulk-Boundary Correspondence}

The two Hamiltonians, Eq.~\eqref{Skyrm-mod1} and Eq.~\eqref{Skyrm-mod2}, each possess a $\mathcal{C}'$-invariant Chern insulator with even Chern number and its TR partner. For Eq.~\eqref{Skyrm-mod1}, the magnitude of this Chern number is $2$, and for Eq.~\eqref{Skyrm-mod2}, it is $4$. For non-negligible SOC or other perturbations breaking $S_z$ conservation, the $\mathbb{Z}_2$ projector invariant calculated from $\nu=(C-C_{TR})/2\  \text{mod}(2)$ will always be trivial for these Chern numbers even as long as the bulk gap does not close according to the ten-fold way~\cite{ten-fold-way,10-fold-way}. One therefore expects the edge states present for zero SOC associated with the Chern insulators will hybridize and gap out immediately for finite coupling between the sectors related by TRS. More generally, all even Chern number Hamiltonians coupled to their TR partners are predicted to yield a topologically-trivial phase due to $
\mathbb{Z}_2$ classification in these cases. 

We observe a counterexample to this statement, that all Hamiltonians with even total Chern number coupled to their TR partners are predicted to yield a topologically-trivial phase, for $\nu_Q$ odd in value, however. We first consider the Hamiltonians (Eq.~\eqref{Skyrm-mod1} and Eq.~\eqref{Skyrm-mod2} for $\lambda=0,c=0$ corresponding to negligible SOC and negligible bulk perturbation, respectively. Each exhibits localized edge states in its slab spectrum, that connect bulk conduction and valence bands. We observe an even number of Kramers pairs at the boundary in agreement with TR symmetry. For a QSHI, finite $c$ corresponding to non-negligible SOC term is expected to gap out the helical edge modes as the topological classification associated with mappings to projectors onto occupied sites reduces from $\mathbb{Z}$ to $\mathbb{Z}_2$. However, in the model considered here, the boundary states appearing from applying OBC to the Hamiltonian \eqref{Skyrm-mod1} do not gap out for finite $c$ and non-negligible SOC. The helical modes instead remain gapless and localized on the edges. Furthermore, the edge states of Eq.\eqref{Skyrm-mod2} gap out for finite $c$ corresponding to non-negligible SOC, with the energy gap being proportional to $c$, as expected for a trivial phase. This shows there is some correspondence between additional robustness of the helical edge modes and $\nu_Q$ odd in value. \\

\begin{figure}[ht!]
    \subfigure[]{\includegraphics[width=0.49\columnwidth,height=3.5cm]{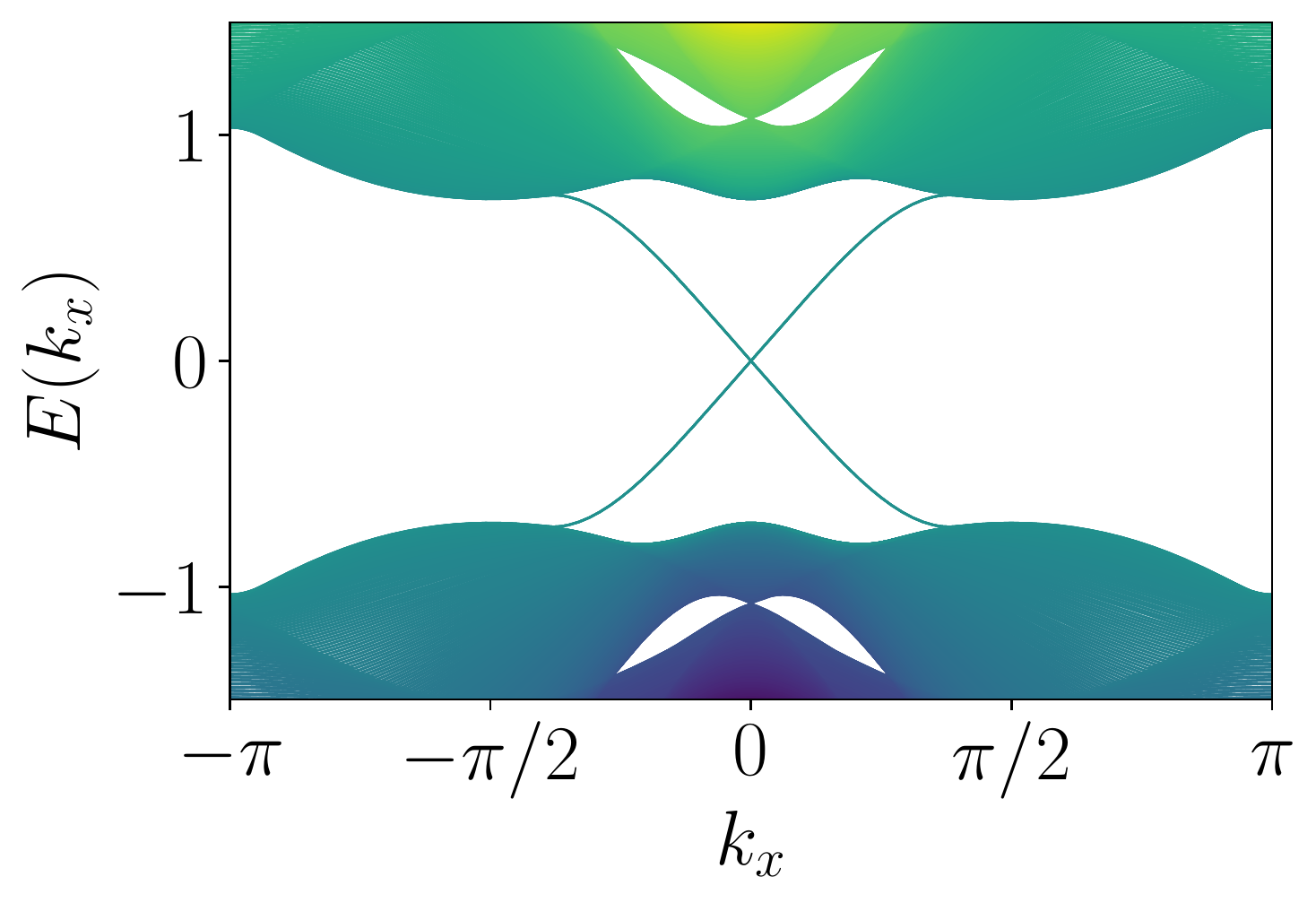}}
    \subfigure[]{\includegraphics[width=0.49\columnwidth,height=3.5cm]{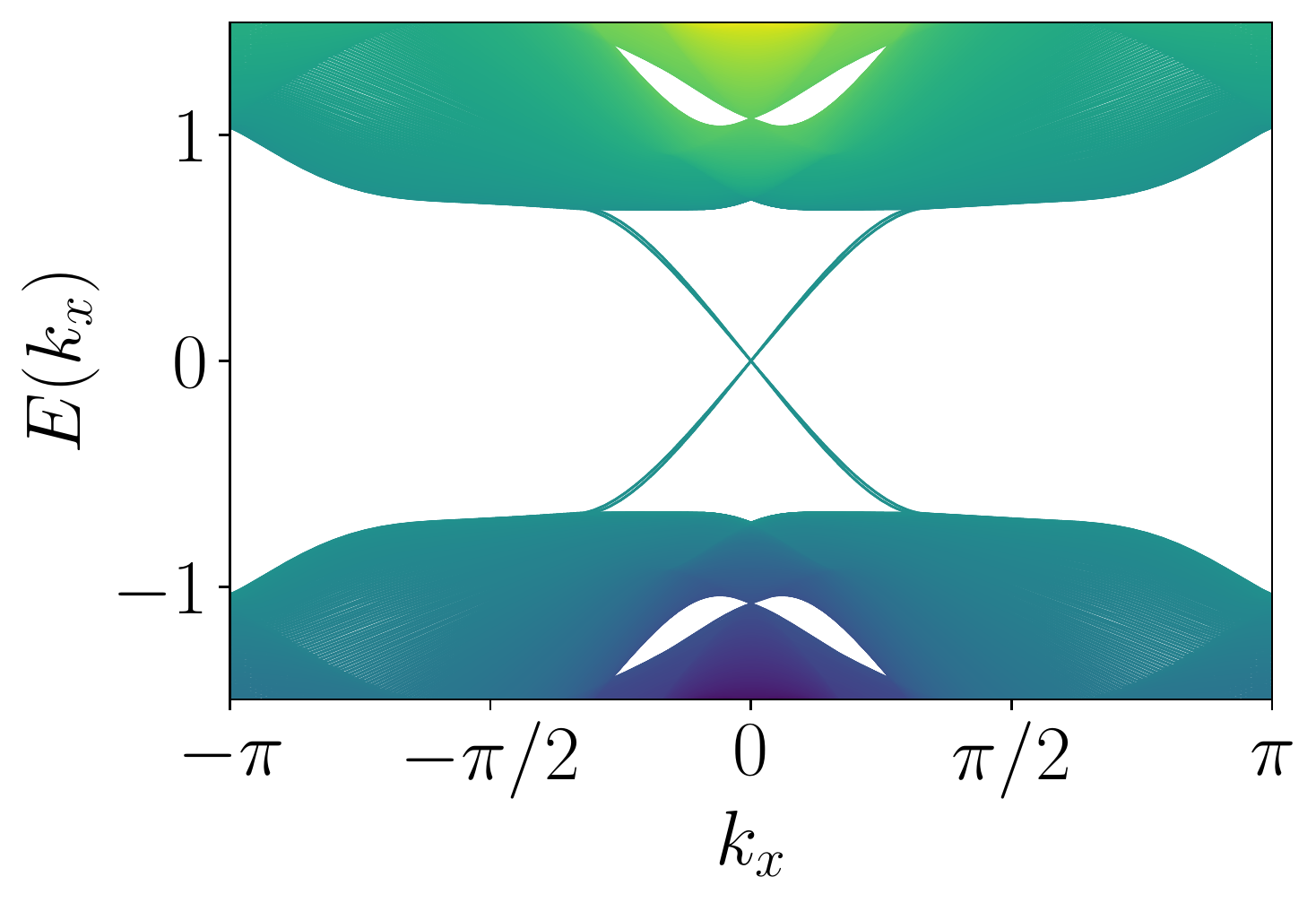}} 
     \subfigure[]{\includegraphics[width=0.49\columnwidth,height=3.5cm]{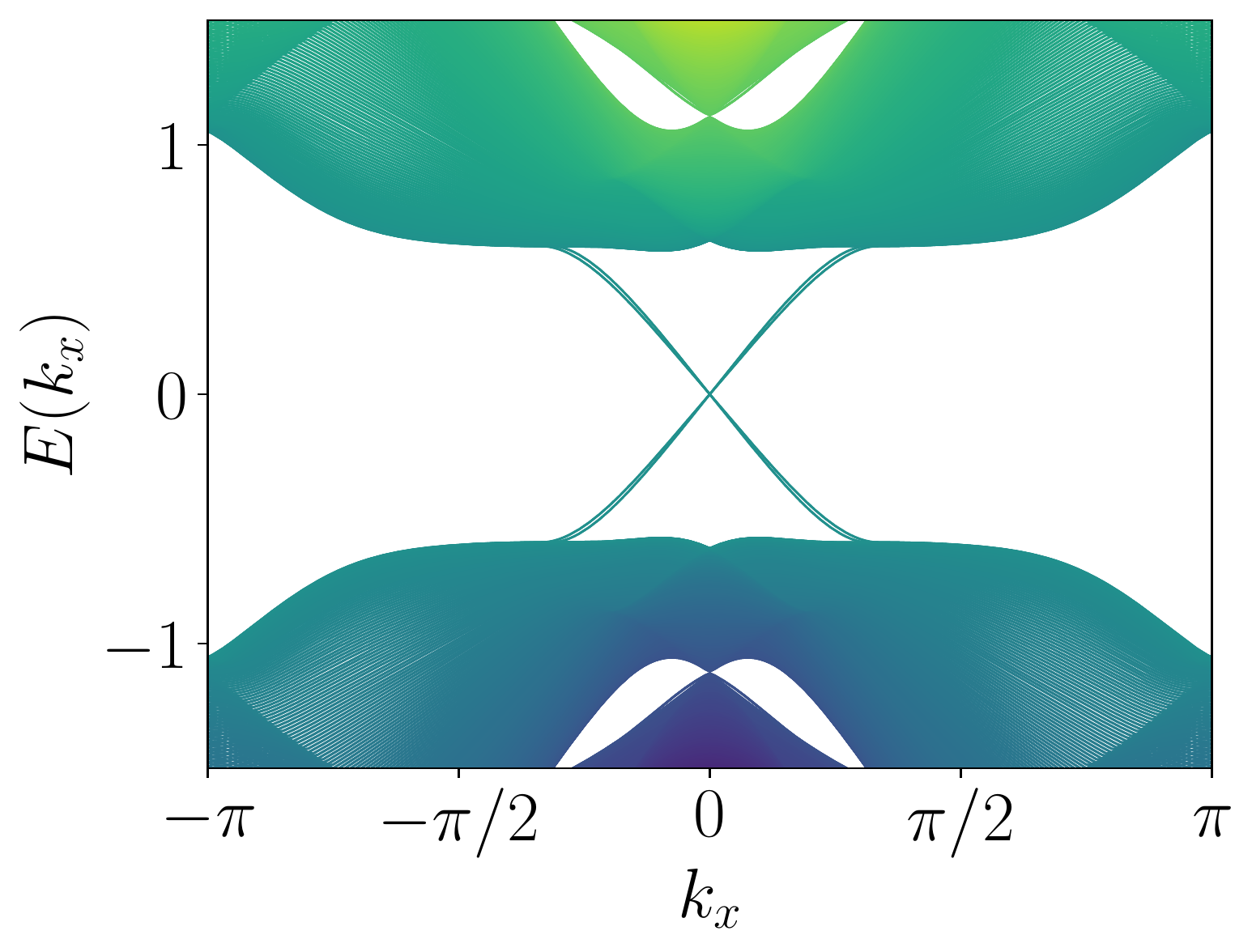}}
    \subfigure[]{\includegraphics[width=0.49\columnwidth,height=3.5cm]{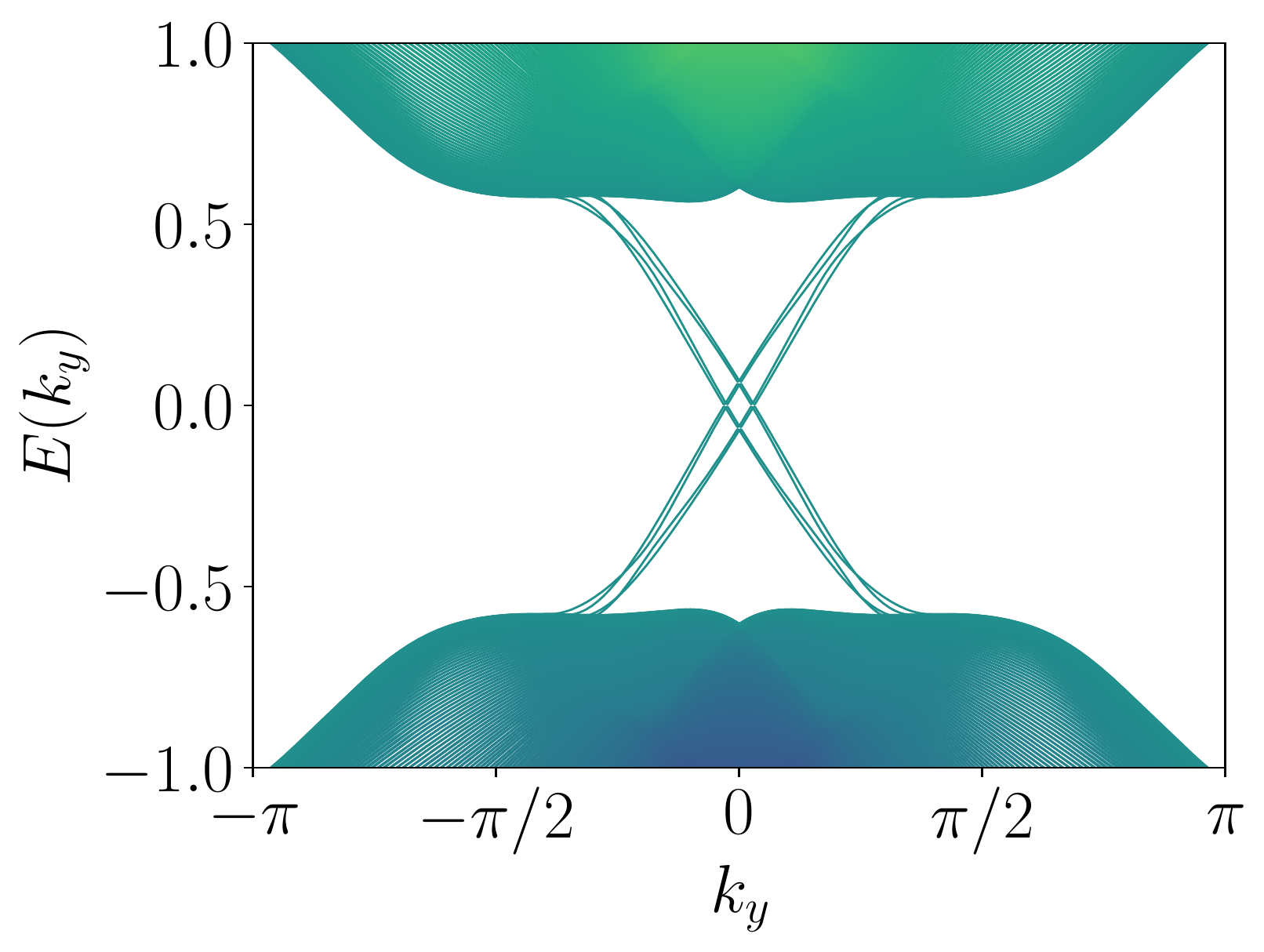}}
  \subfigure[]{\includegraphics[width=0.49\columnwidth,height=3.5cm]{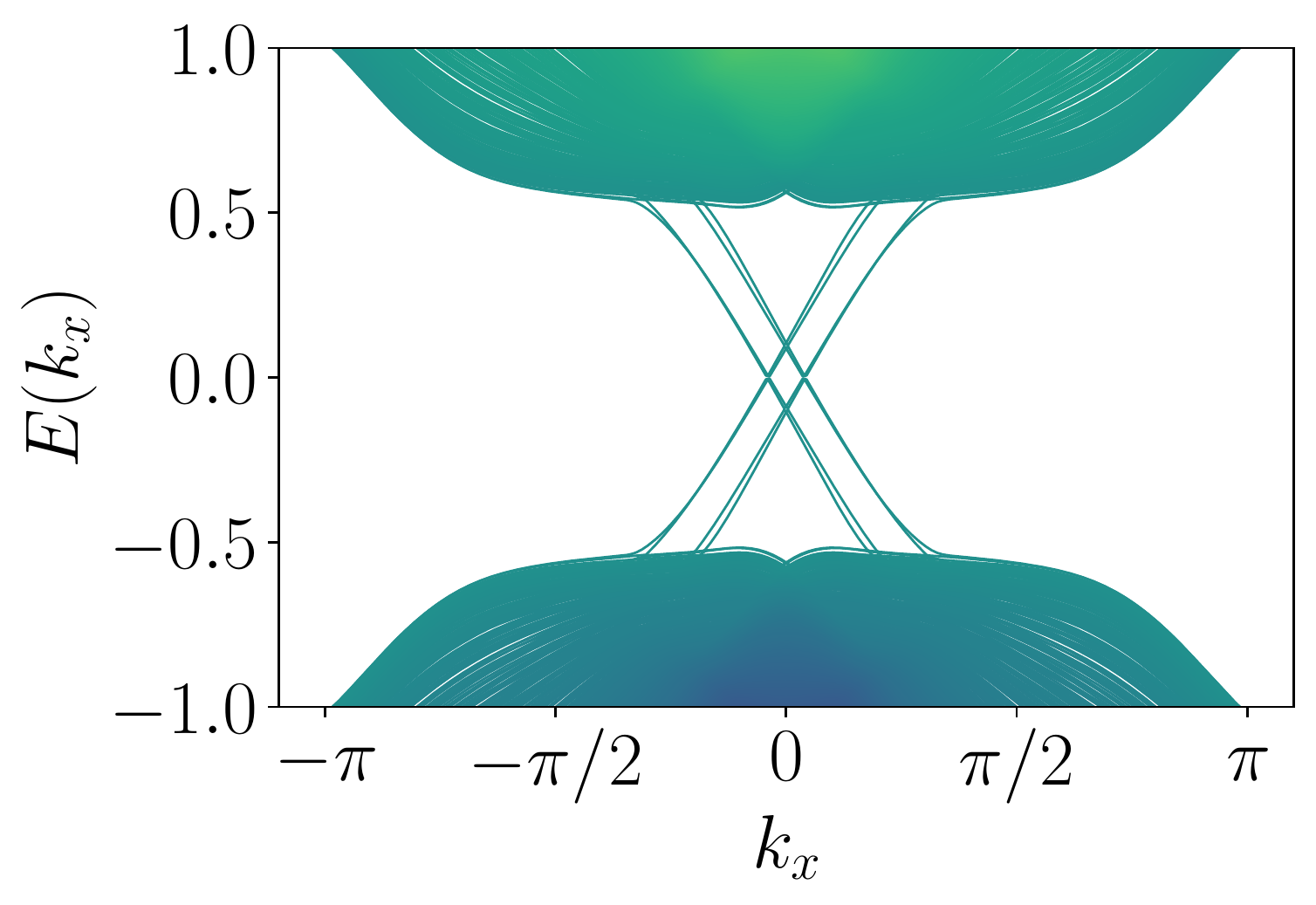}}
  \subfigure[]{\includegraphics[width=0.49\columnwidth,height=3.5cm]{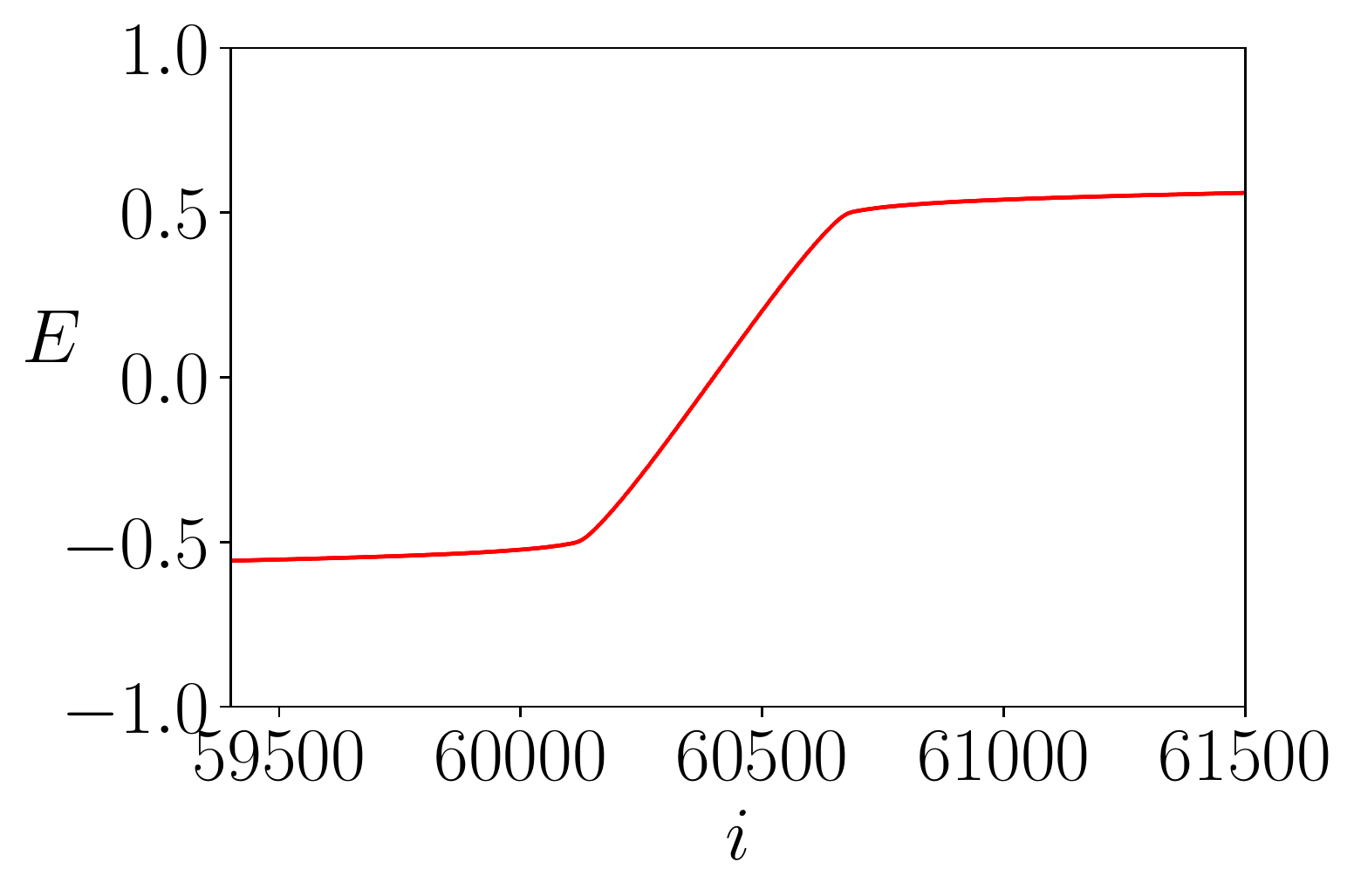}}
  \vspace{-0.3cm}
  \caption{a) Energy spectrum for the model of Eq. \eqref{Skyrm-mod1} with OBC in $y$ and periodic in $x$ with parameters $M=-1$, $\Delta_0 = 0.5$,$c=0.4$ length $N_y=200$ and $\lambda=0.0$. b) Energy spectrum for the same conditions except $\lambda=0.3$. c) Edge spectrum with OBC in $y$ for $M=-1$, $\Delta_0 = 0.5$,$c=0.5$,$\lambda=0.3$ with the edge perturbation discussed in the text $V_1=0.1$. d) Same parameters and perturbation but now with OBC in $x$. e) Edge spectrum with OBC in $y$ for the same parameters as c) but with some random disorder realization respecting particle-hole and TR. d) Same parameters and perturbation but now with OBC in $x$. f) Energy spectrum as a function of eigenvalue index $i$ for a disorder average over $100$ on-site potentials $V_j s_0\tau_z\sigma_0$ distributed with zero mean and standard deviation $0.1$. For a system with open boundary conditions in $y$, periodic in $x$ and parameters $M=-1$,$\Delta_0=0.5$,$c=0.5$, $\lambda=0.3$. }
  \label{BBC_1}
\end{figure}

We further investigate the mechanism for this unexpected robustness of the helical edge modes for trivial $\mathbb{Z}_2$ projector invariant and non-trivial $\mathbb{Z}_2$ skyrmion invariant by considering $\lambda$ finite, where $\lambda$ is the free parameter of a bulk perturbation that breaks all symmetries of Eq.~\eqref{Skyrm-mod1}, except for TR and particle-hole. The results of this perturbation are summarized in Fig. \ref{BBC_1}. Fig.~\ref{BBC_1} a) and b) show that, even with all crystalline symmetries broken, the gapless edge states are present. This shows that the gapless edge states are not protected by crystalline point group symmetries of the bulk.\\

We investigate the robustness of the gapless modes against edge perturbations. Fig. \ref{BBC_1} c) and d) show the slab spectra for an additional edge perturbation in Eq. \eqref{Skyrm-mod1}, which has previously been shown to gap edge states of the QSHI~\cite{Doru-Sticlet}:
\begin{align}
    V_{\text{edge}}(y,k_x)=\begin{cases}
    V_1\cos(k_x)s_0\tau_0\sigma_z & y=0 \\
    -V_1\sin(k_x)s_x \tau_z\sigma_0 & y=N_y\\
    0 & \text{else}
    \end{cases}
\end{align}
This perturbation can also be applied for the case of OBC in the $x$ direction by interchanging the $x$ and $y$ labels in its definition. As is clear from the plots in Fig.~\ref{BBC_1} a)-d) the gapless points may shift, but they remain present even in the present of spin-orbit coupling. 

We also investigate the effects of on-site disorder on the edge dispersion and localization of the edge states protected by the skyrmion topology. We add on-site disorder of the form $V_j s_0\tau_z\sigma_0$ for OBC in $y$, with $j$ being the layer index. This form of disorder preserves time-reversal and particle-hole symmetries. The effects on the energy spectrum are shown in Fig.~\ref{BBC_1} e) and f), respectively. We find that the gapless edge states persist in the presence of a random disorder realization, as shown in Fig.~\ref{BBC_1} e). To compute corresponding disorder-averaged results, we compute the spectrum of Eq.~\eqref{Skyrm-mod1} in this slab geometry for each $k_x$, and then combine these energy eigenvalues for each $k_x$ into a single larger array, which we then sort in energy. This is computed for each of 100 disorder realizations, and the average of these sorted spectra is shown in Fig.~\ref{BBC_1} f). We find the edge states of this disorder-averaged spectrum remain gapless, corresponding to each disorder realization simply shifting the crossing point of the helical modes in $k_x$.\\

\begin{figure}[ht!]
    \subfigure[]{\includegraphics[width=0.49\columnwidth,height=3.3cm]{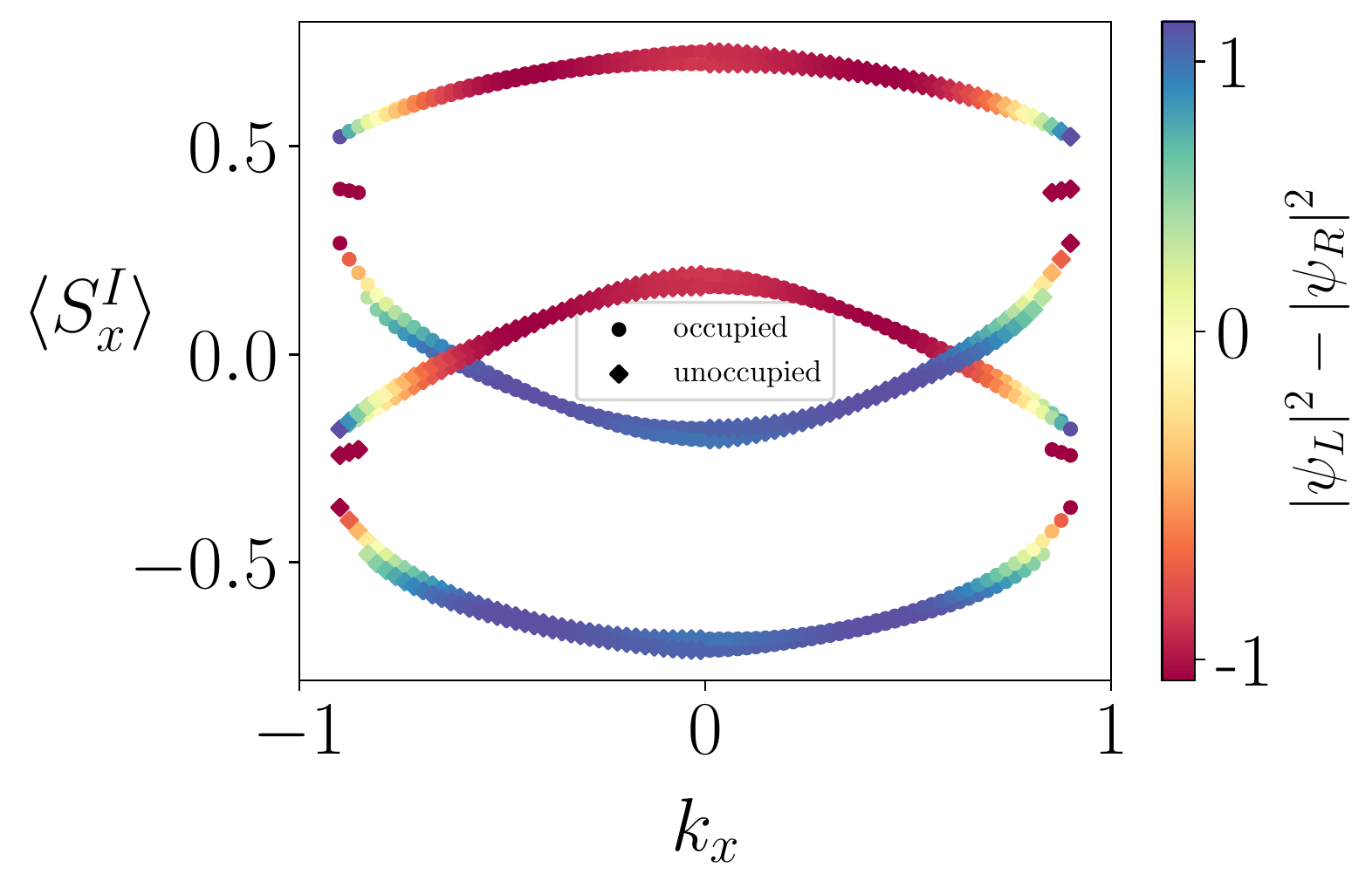}}
    \subfigure[]{\includegraphics[width=0.49\columnwidth,height=3.3cm]{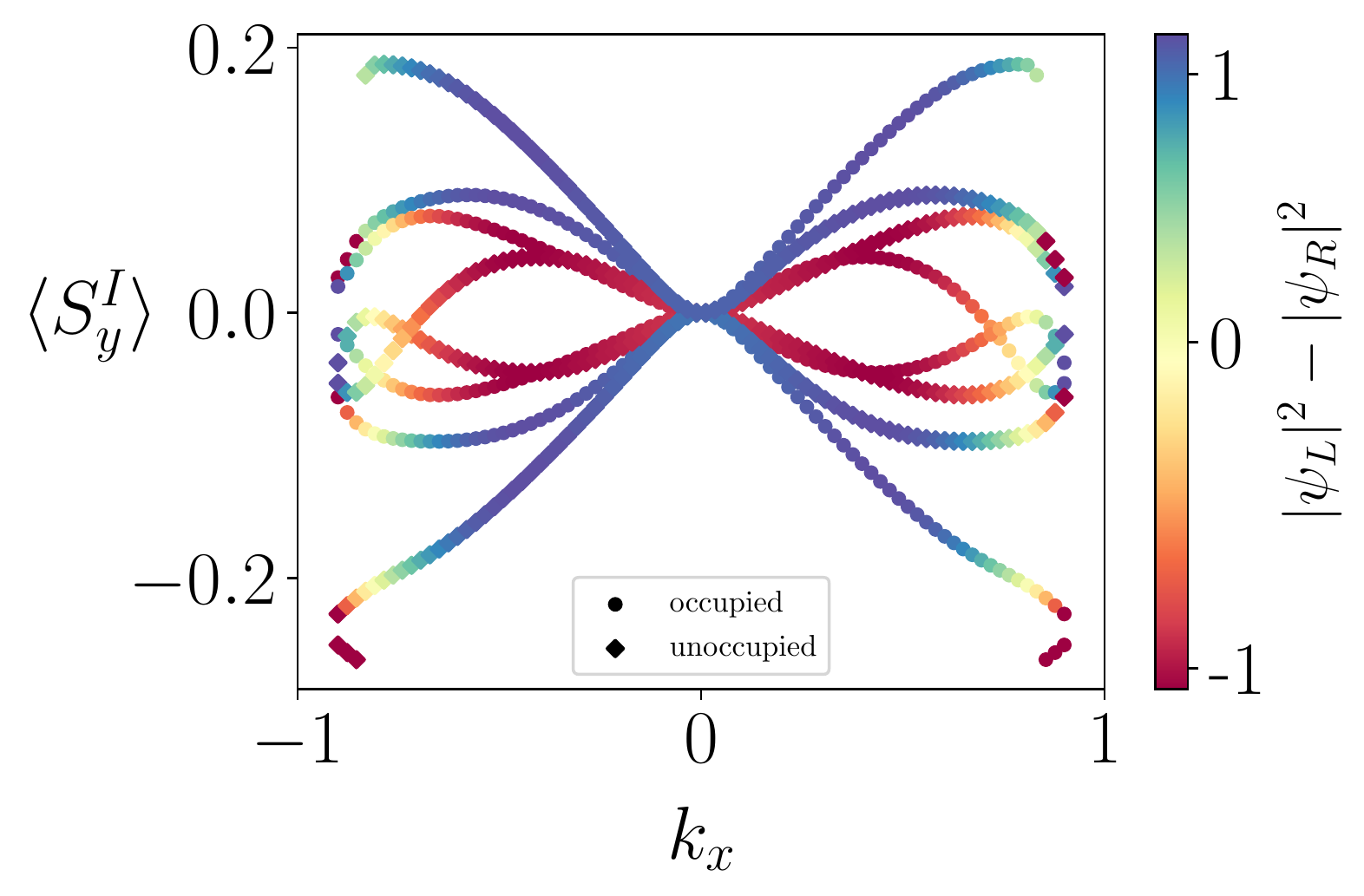}} 
     \subfigure[]{\includegraphics[width=0.49\columnwidth,height=3.3cm]{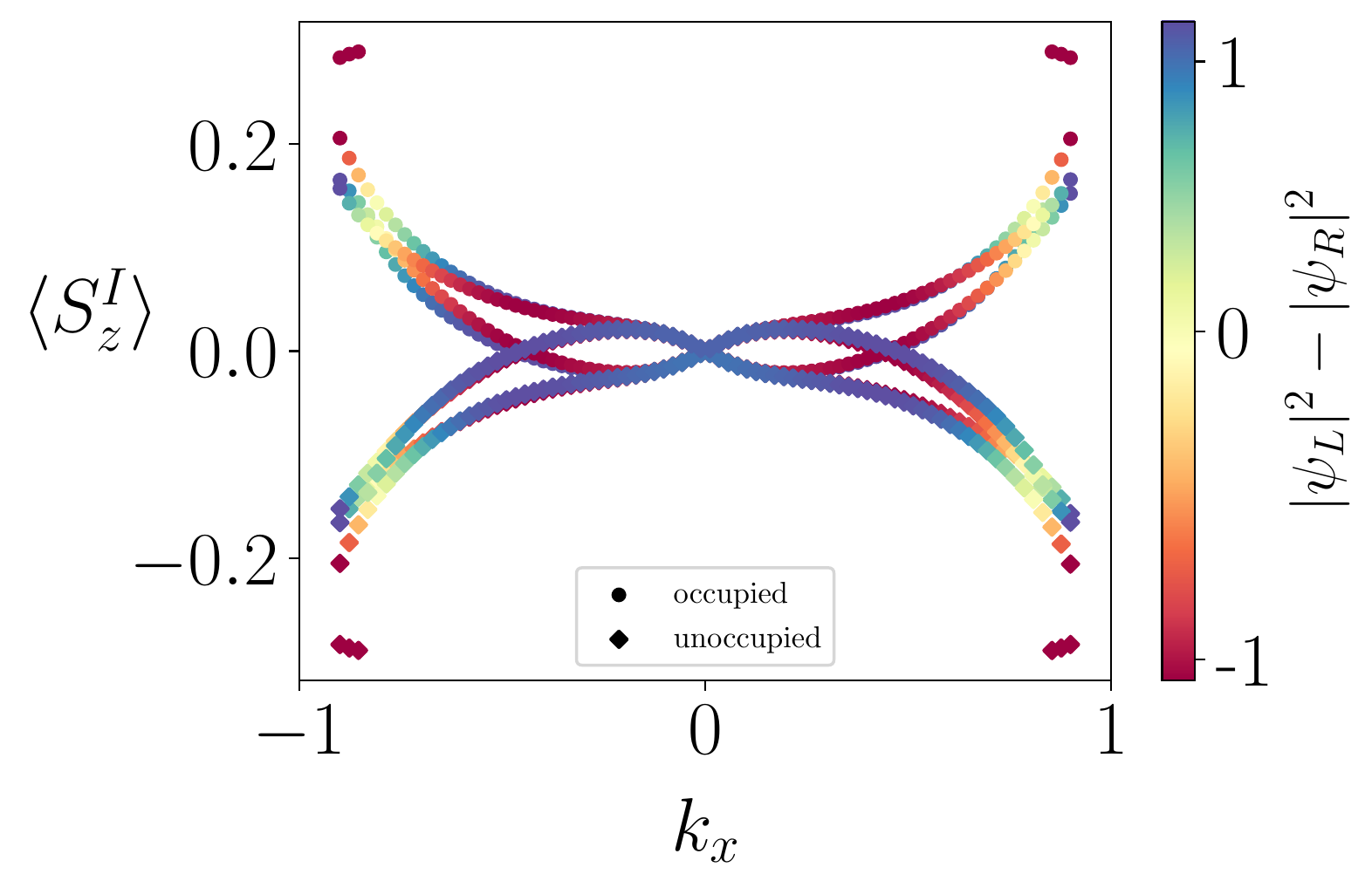}}
    \subfigure[]{\includegraphics[width=0.49\columnwidth,height=3.3cm]{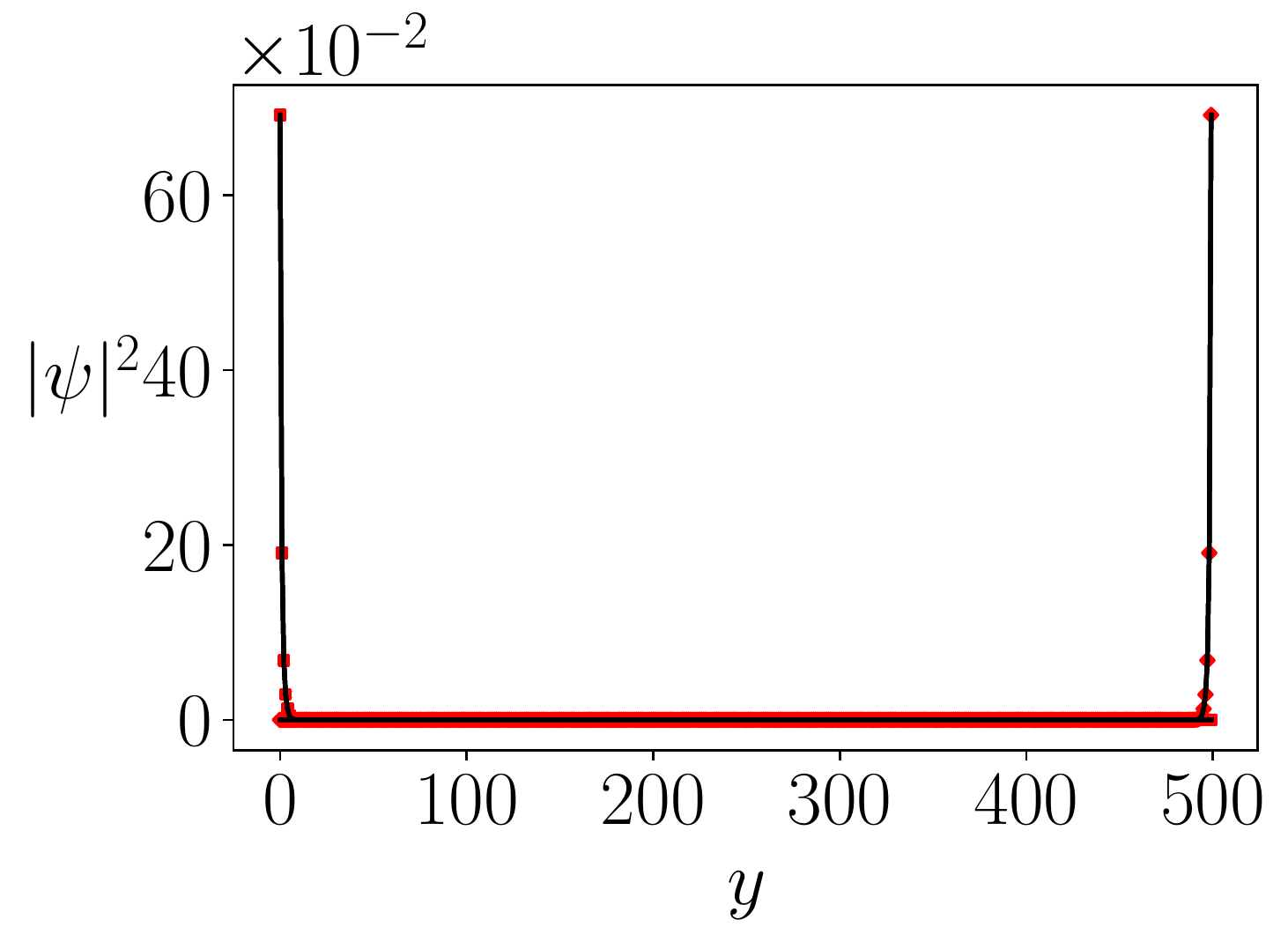}}
    \subfigure[]{\includegraphics[width=0.49\columnwidth,height=3.3cm]{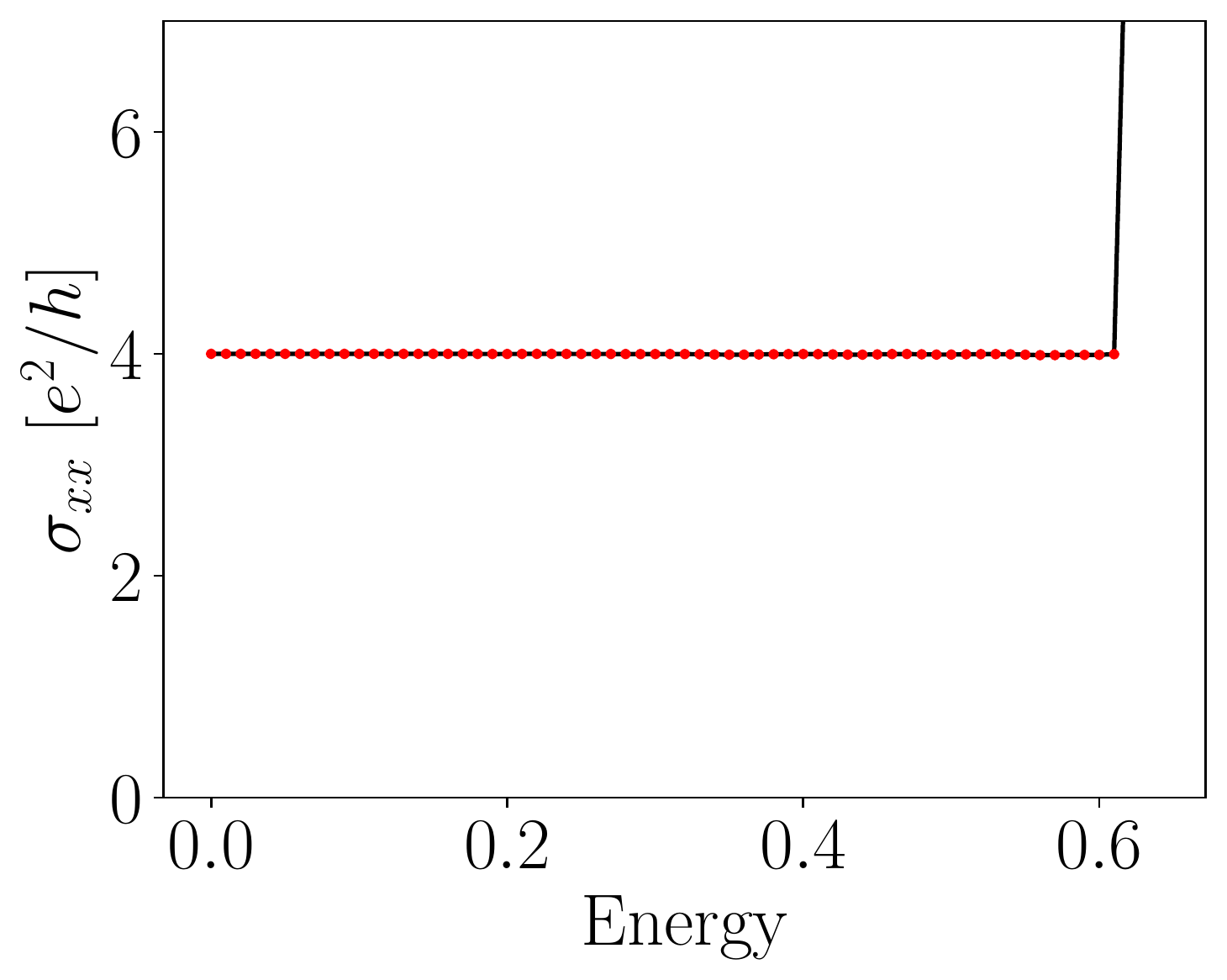}}
    \subfigure[]{\includegraphics[width=0.49\columnwidth,height=3.3cm]{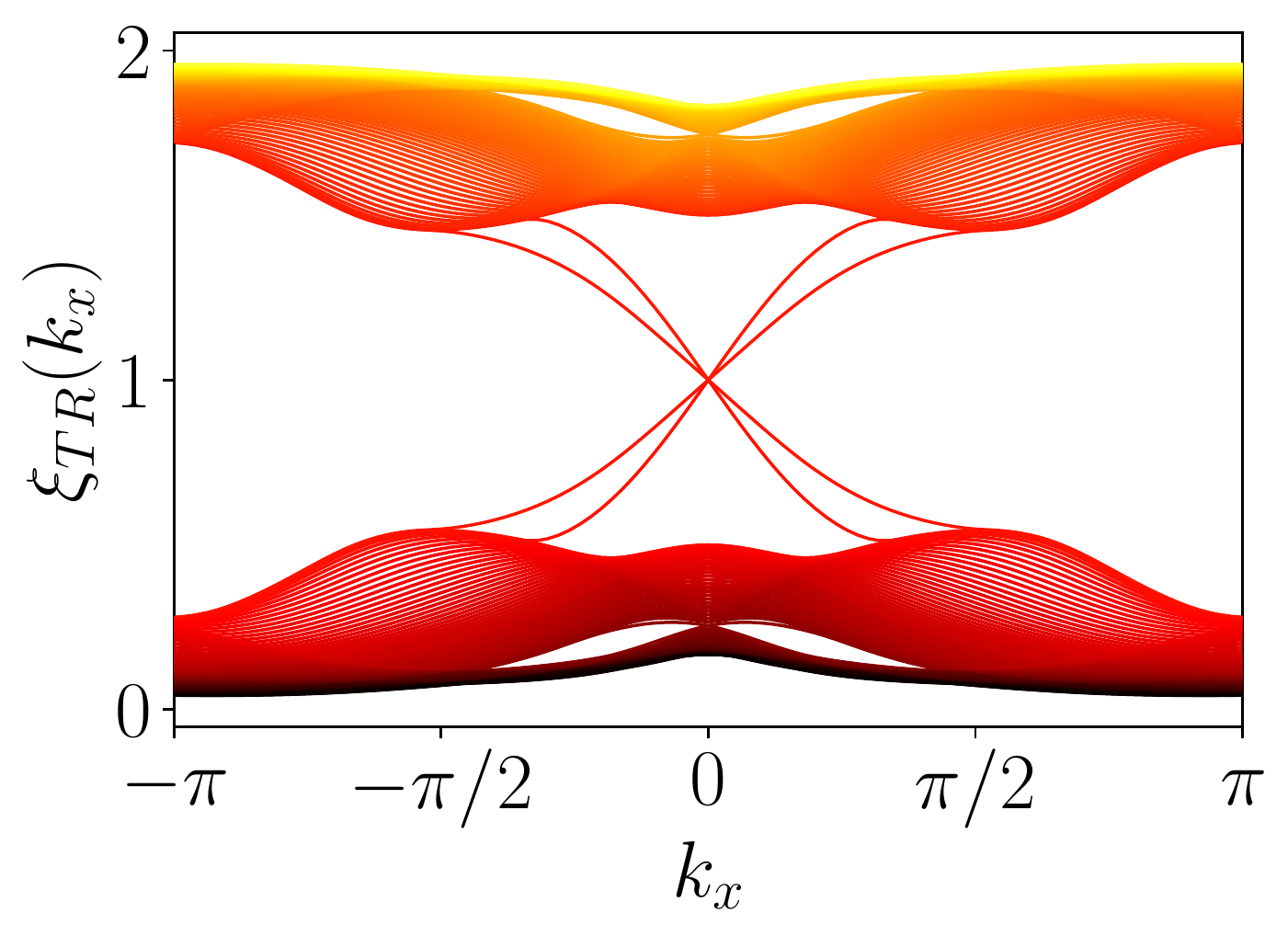}}
  \vspace{-0.3cm}
  \caption{a) $x$ component of the pseudo-spin expectation value for the edge states of the Hamiltonian in equation \ref{Skyrm-mod1} with parameters $M = -1$,$\Delta_0 = 0.5$, $\lambda = 0.3$ , $c = 0.5$ and open boundary conditions on $y$. The color is proportional to the localization on either edge computed by taking the probability density at each edge and subtracting them. b) and c) show the $y,z$ components of the pseudo-spin expectation value for the edge states respectively. d) Probability density for the edge states of the Hamiltonian in equation \ref{Skyrm-mod1}.e) Conductivity along $xx$ as a function of energy for the Hamiltonian in equation \ref{Skyrm-mod1} with parameters $M = -1$,$\Delta_0 = 0.5$, $\lambda = 0.0$ , $c = 0.5$ and full open boundary conditions, random gaussian disorder along the whole sample with standard deviation $0.01$ is present.f) 2-point reduced correlation spectrum for the Hamiltonian in equation \ref{Skyrm-mod1} with parameters $M = -1$,$\Delta_0 = 0.5$, $\lambda = 0.3$ , $c = 0.5$ and all $\mathcal{T},\mathcal{C}$ symmetry allowed perturbations with amplitude $0.5$ turned on, for a cut along the $y$ axis and traced over $L=200$ lattice sites and particle-hole d.o.f.}
  \label{Spin-polariz-edge}
\end{figure}

We now also explore how the non-trivial skyrmion invariant $\nu_{\mathcal{Q}}$ alters the character of the edge states to prevent hybridisation. Given related work introducing the observable-enriched partial trace and characterizing the additional spin-momentum-locking of edge states due to non-trivial skyrmion number~\cite{winter2022}, we compute the pseudo-spin expectation values for the edge states for a given TR sector in a slab geometry with either $x$ or $y$ directions open so as to have $k_y$ or $k_x$ as a good quantum number. We find that, even though non-neglible SOC is present, the edge state pseudo-spin texture has a very distinct pattern depending on the orientation of the edge. For open boundary conditions in the $y$ direction, near the crossing of the edge modes the pseudo-spin $\expval{S_y^I},\expval{S_z^I}$ seems to average to zero for each edge, while $\expval{S_x^I}$ does not as seen in Fig.\ref{Spin-polariz-edge}. Instead the value of $\expval{S_x^I}$ on each edge seems to be minus the one of the other edge.\\

We additionally confirm the nontrivial nature of the edge states by computing the $\sigma_{xx}$ conductivity as a function of energy as shown in Fig. \ref{Spin-polariz-edge} e). Here the conduction band of the system starts around $0.6$ in units of the hopping, since this was set to be one by default, and so a trivial zero conductance would be expected below it. Instead a quantized longitudinal conductivity of $4e^2/h$ appears, even in the presence of disorder. Such a signature implies the existance of transmision channels present in the system which necessarily come from edge state transmission since the bulk material is an insulator \cite{Proskurin_2015,AbrikosovConduct,GorbarConduct}. Specifically for the case of the QSHI it has been shown that a quantized transport robust to disorder, as in Fig.\ref{Spin-polariz-edge} e), is indeed linked to the presence of topologically protected edge states which for the QSHI carry a $2e^2/h$ conductance \cite{QTransportQSHI}. We indeed observe a quantization to $4e^2/h$ consistent with the four edge states per edge which appear when opening boundary conditions as discussed in the previous paragraphs. The robustness to disorder for this $\nu=0,\nu_{\mathcal{Q}}= 1$ case is consistent with quenching of SOC by the topological spin polarization of the edge states that is part of the bulk-boundary correspondence of topological skyrmion phases of matter. We may also understand it from the perspective of the lower-symmetry realizations of topological skyrmion phases in study of the quantum skyrmion Hall effect~\cite{QSkHE}: three-band tight-binding models for topological skyrmion phases of matter without $\mc{C}'$ symmetry realize generalized Thouless pumps, in which spin angular momentum is pumped, rather than electric charge. The spin polarization of edge states observed here corresponds to a $\mc{C}'$-symmetric and TR-symmetric Thouless pump of spin angular momentum, and gaplessness of edge states is required for this pumping in correspondence with $\nu_{Q}$ non-trivial.\\

However, the most robust astonishing feature associated to the skyrmion invariant is seen upon tracing out the particle-hole degree of freedom in each spin sector, by performing the observable-enriched partial trace defined in Appendix S3. That is, we characterize the skyrmion topology by tracing out the particle-hole degree of freedom---in addition to a virtual cut in real-space---to compute a reduced two-point function and corresponding entanglement spectrum computed from the ground state as seen in Fig. \ref{Spin-polariz-edge} f) . In correspondence with the non-trivial skyrmion invariant, this observable-enriched entanglement spectrum exhibits helical edge modes. We may therefore understand the spin polarization of edge states in the full system as consequences of the bulk-boundary correspondence of a potentially open subsystem, realized by tracing out the particle-hole degree of freedom.

In addition to the previous characterization of the edge states, we finally also explore the remarkable robustness of the gapless helical edge modes for trivial projector invariant and non-trivial skyrmion invariant, by enumerating all possible spin-orbit coupling terms that preserve time-reversal symmetry and particle-hole symmetry. Out of all possible six momentum-even terms, the gapless helical edge modes are robust against three of these terms, and the other three gap them out. These terms are provided in appendix S4. We distinguish the three SOC terms preserving the gapless helical edge modes from those which gap them out by additional symmetry-protection. 

First, we find that certain $4\times4$ submatrices of the Hamiltonian must preserve $\left(\mathcal{C} \mathcal{T}\right)_p$ symmetry (defined in appendix S4), otherwise an effective Rashba SOC term is introduced, permitting finite $y$ component of spin in sector I/II for the edge states in the vicinity of the gapless point(s) in the edge spectrum. That is, the spin-momentum-locking of the edge states is relaxed. As discussed in appendix S4, clockwise spin rotation about the $z$-axis within sector I/II for one state at the edge is compensated by counter-clockwise rotation of the second state at that edge within the same sector (the $\mc{C}'$ partner of the first state), such that the net spin of these edge states in combination remains polarized in the $x$-direction. We may compute the observable-enriched entanglement spectrum as defined in Appendix S3 to examine bulk-boundary correspondence in our system upon tracing out the particle-hole degree of freedom of each spin sector to further characterize this edge state spin texture, shown in Appendix S4. We find the spin textures are consistent with gapless states of the spin subsystem propagating along the edge deep within the bulk gap of the entanglement spectrum. Indeed, we examine the edge spectrum of the spin subsystem explicitly to show the non-trivial spin topology is preserved in this lower-symmetry case, but the observable-enriched entanglement spectrum exhibits additional $k_x$-dependent Rashba splitting. This finite $y$ spin component permits hybridisation between edge states previously forced to zero by the stricter spin-momentum-locking for $\left(\mc{C} \mc{T}\right)_p$ preserved, gapping out the helical edge states protected by the spin topology.

Second, in this system, we have three two-fold degrees of freedom, corresponding to the $s$, $\tau$, and $\sigma$ Pauli matrices, respectively. The symmetry operator $\mathcal{C}'$ , which acts on the $\tau$ and $\sigma$ degrees of freedom only ( acting simply as a constant in the $s$ sector), is required to define two skyrmion numbers, $\mathcal{Q}_I$ and $\mathcal{Q}_{II}$, computed as winding of spin $\mathcal{S}_I$ and $\mathcal{S}_{II}$ in terms of the matrices discussed in appendix S2, respectively. The precise statements of these symmetry requirements are provided in appendix S4. This additional symmetry protection reflects the fact that the helical edge modes are being protected by topology of subsets of the degrees of freedom, and indicates the concept of symmetry-protection for topological phases must be generalized in light of the topological skyrmion phases of matter.\\

\section{Conclusions}

In this paper, we introduce time-reversal-invariant topological skyrmion phases of matter. These are topological phases realized in systems with time-reversal symmetry as required to protect the quantum spin Hall insulator phase, but which are distinct from the QSHI phase. The QSHI is a topological phase of the ten-fold way classification scheme, resulting from mappings from the full Brillouin zone to the space of projectors onto occupied states, while the TRI skyrmion phase arises from topologically non-trivial mappings from the full Brillouin zone to the space of ground state spin expectation values. The TRI skyrmion phases are therefore characterized in the bulk by formation of a skyrmion in the texture of the ground-state spin expectation value over the Brillouin zone in general, rather than in the texture of the projector onto occupied states over the Brillouin zone.\\

We construct toy models for the TRI skyrmion phase by combining the Bloch Hamiltonian for a $\mathcal{C}'$-invariant but time-reversal symmetry-breaking topological skyrmion phase with its time-reversed partner into a Bloch Hamiltonian matrix representation with time-reversal symmetry. We may therefore characterize the TRS skyrmion phase with two skyrmion numbers related to one another by TRS in simpler cases corresponding to a helical topological skyrmion phase. More generally, we may also included terms $V_{\mathrm{SOC}}$ coupling these two sectors, in which case the phase is characterized by a single topological invariant $\nu_{\mathcal{Q}}$, which is equal to the skyrmion number for one sector, modulo $2$. This corresponds to a more general $\mathbb{Z}_2$ classification of $\nu_{\mathcal{Q}}$.\\ 

Performing an observable-enriched partial trace on the density matrix of the occupied states for a system with open boundary conditions corresponding to a slab geometry to compute the observable-enriched reduced density matrix of the spin subsystem, we find odd $\nu_{\mathcal{Q}}$ corresponds to topologically-protected gapless modes in the observable-enriched entanglement spectrum. We show this bulk-boundary correspondence of the spin subsystem has consequences for the bulk-boundary correspondence of the full system, demanding robust helical gapless edge states at the boundary of the system for open boundary conditions, even when the system is not in a quantum spin Hall insulator phase according to the $\mathbb{Z}_2$ projector invariant, $\nu$. That is, for $\nu$ even and trivial, but $\nu_{\mathcal{Q}}$ odd and non-trivial, we find topologically-robust, helical gapless edge modes protected by non-trivial $\nu_{\mathcal{Q}}$. When each of $\nu$ and $\nu_{\mathcal{Q}}$ is even, topologically-protected helical edge modes are absent.\\

We find the persistence of the helical modes for $\nu_{\mathcal{Q}}$ non-trivial---when $\nu$ is trivial---is related to the strict spin-momentum-locking enforced on edge states in the full system by $\nu_{\mathcal{Q}}$ non-trivial. This spin-momentum-locking derives from the requirement of gapless helical boundary modes in the observable-enriched slab entanglement spectrum of the spin subsystem due to non-trivial $\nu_{\mathcal{Q}}$ in the bulk. We add terms to the bulk Hamiltonian such that projectors onto the occupied states possess only time-reversal symmetry $\mathcal{T}$ and particle-hole symmetry $\mathcal{C}$. We find that the helical gapless modes observed for $\nu$ trivial and $\nu_{\mathcal{Q}}$ non-trivial are robust for non-negligible $V_{\mathrm{SOC}}$ respecting a $\mathcal{C}'$ symmetry of a subset of the degrees of freedom, as well as lacking a generalized Rashba SOC, which softens spin-momentum-locking at the edge while still yielding the gapless helical edge modes of the spin subsystem after observable-enriched partial trace. Future work will investigate the potential of TRI six-band models constructed from the lower-symmetry three-band models for topological skyrmion phases~\cite{QSkHE} to realize topologically-robust gapless boundary modes, with the expectation that absence of $\mc{C}'$ symmetry in these models will yield more robust helical edge modes than those observed in the $\mc{C}'$ symmetric cases discussed here, as well as type-II topological phase transitions~\cite{QSkHE, First-skyrmion}.  

Notably, these helical gapless modes affect computation of the projector topological invariant: spectral flow winds non-trivially in correspondence with presence of the helical gapless modes, contradicting $\mathbb{Z}_2$ classification of $\nu$. This indicates the projector invariant $\nu$ must be generalized to incorporate changes due to $\nu_{\mathcal{Q}}$, rather than treated as independent of $\nu_{\mathcal{Q}}$. This is also consistent with results of three-band models for topological skyrmion phases, where non-trivial total Chern number as determined by bulk spectral flow does not necessarily yield chiral transport on the edge. These failures relate to reliance of much topological characterization on the flat-band limit assumption and topological stability up to closing of a charge gap, which does not hold in general~\cite{First-skyrmion, QSkHE}.

Such topologically-protected gapless helical modes and additional spin-momentum-locking may be observed in transport measurements or spin-ARPES and serve as important signatures and consequences of topological skyrmion phases of matter. The realization of these signatures in time-reversal invariant systems is also highly-desirable for experiment, as the non-trivial topology is realized without the need for particular magnetic orders which generally complicate experimental realization. Our work is therefore significant in bringing study of topological skyrmion phases---including their role in extending topological classification schemes---much closer to experimental realization.

\textit{Acknowledgements}---This research was supported in part by the National Science Foundation under Grant No. NSF PHY-1748958.

\bibliography{TR-Skyrmion}

\clearpage

%%%%%%%%% Prefix a "S" to all equations, figures, tables and reset the counter %%%%%%%%%%
\makeatletter
\renewcommand{\theequation}{S\arabic{equation}}
\renewcommand{\thefigure}{S\arabic{figure}}
\renewcommand{\thesection}{S\arabic{section}}
\setcounter{equation}{0}
\setcounter{section}{0}
\onecolumngrid
\begin{center}
  \textbf{\large Supplemental material for ``Time-reversal invariant topological skyrmion phases''}\\[.2cm]
  R. Flores-Calderon,$^{1,2}$ and Ashley M. Cook$^{1,2,*}$\\[.1cm]
  {\itshape ${}^1$Max Planck Institute for Chemical Physics of Solids, Nöthnitzer Strasse 40, 01187 Dresden, Germany\\
  ${}^2$Max Planck Institute for the Physics of Complex Systems, Nöthnitzer Strasse 38, 01187 Dresden, Germany\\}
  ${}^*$Electronic address: cooka@pks.mpg.de\\
(Dated: \today)\\[1cm]
\end{center}
\section{Proof that the total   skyrmion number is zero for TR systems }

We start with the expression for the total skyrmion number introduced in ref. \cite{First-skyrmion}:
\begin{align}
    \mathcal{Q}=\frac{1}{4 \pi} \int_{\text{BZ}} \Omega_{\hat{\mathcal{S}}}(\mathbf{k}) d^{2} k
\end{align}
where $\Omega_{\hat{\mathcal{S}}}(\mathbf{k})=\hat{\mathcal{S}}(\mathbf{k}) \cdot\left(\partial_{k_{x}} \hat{\mathcal{S}}(\mathbf{k}) \times \partial_{k_{y}} \hat{\mathcal{S}}(\mathbf{k})\right)$ and we have the normalized spin ground state texture $\hat{\mathcal{S}}(\mathbf{k})=\mathcal{S}(\mathbf{k}) /|\mathcal{S}(\mathbf{k})|$. We now want to consider a TR system and the effect of reversing momentum on the GS spin expectation value, meaning calculating $\mathcal{S}(-\mathbf{k})$:
\begin{align}
    \mathcal{S}(-\mathbf{k})_\mu= \expval{S_\mu}_{-\mathbf{k}}=\sum_{n=1}^{n_{\text{occ}}} \expval{S_\mu}{u_n(-\mathbf{k})}=\sum_{m=1}^{n_{\text{occ}}/2} \expval{S_\mu}{u_{m I}(-\mathbf{k})}+\expval{S_\mu}{u_{m II}(-\mathbf{k})}
\end{align}
where we have expressed the ground state expectation value in terms of the energy eigenfunctions $\ket{u_n(\mathbf{k})}$. We also have used TR symmetry by renaming the eigenfunctions such that we separate between the expectation value in the two components of a Kramers pair $I$ and $II$ and relegated all other degrees of freedom to the index $m$ running up to $n_{\text{occ}}/2$. Now we use the important TR relationship between Kramer pairs: 
\begin{align}
    \ket{u_{m I}(-\mathbf{k})}=e^{i \chi_{m \mathbf{k}}} T \ket{u_{m II}(\mathbf{k})}\label{TR-relation}
\end{align}
Where $T$ is the time reversal operator and $\chi_{\mathbf{k}}$ is a real function whose choice determines a gauge choice for the wave functions. Now let us use the anti-unitarity of the time reversal operator to express it as $T= U_t K$ where $U_t$ is a unitary matrix, and $K$ is the complex conjugate operation. Because of spinful TR we know that $U_t=-U_t^T$ so that the previous equation in component form is just 
\begin{align}
    u_{m I}^\alpha(-\mathbf{k})= e^{i \chi_{m\mathbf{k}}} U_t^{\alpha \alpha'} (u_{m II}^{\alpha'}(\mathbf{k}))^*
\end{align}
,where greek indices run over all components and implicit Einstein summation is assumed. Then we obtain for the expectation value of the spin:
\begin{align}
    &\mathcal{S}(-\mathbf{k})_\mu= \expval{S_\mu}_{-\mathbf{k}}=\sum_{m=1}^{n_{\text{occ}}/2} u_{m II}^{\beta'}(\mathbf{k}) (U_t^\dagger)^{\beta' \beta} S_\mu^{\beta \alpha}U_t^{\alpha \alpha'} (u_{m II}^{\alpha'}(\mathbf{k}))^*+u_{m I}^{\beta'}(\mathbf{k}) (U_t^\dagger)^{\beta' \beta} S_\mu^{\beta \alpha}U_t^{\alpha \alpha'} (u_{m I}^{\alpha'}(\mathbf{k}))^*\\
    &=\sum_{m=1}^{n_{\text{occ}}/2} \expval{(U_t^*)^\dagger S^*_\mu U_t^*}{u_{m I}(\mathbf{k})}^*+\expval{(U_t^*)^\dagger S^*_\mu U_t^*}{u_{m II}(\mathbf{k})}^*=\sum_{n=1}^{n_{\text{occ}}} \expval{(U_t^T) S^*_\mu U_t^*}{u_n(\mathbf{k})}^*\\
    &=\sum_{n=1}^{n_{\text{occ}}} \expval{(-U_t) S^*_\mu (-U_t^\dagger)}{u_n(\mathbf{k})}^*=\sum_{n=1}^{n_{\text{occ}}} \expval{U_t S^*_\mu U_t^\dagger}{u_n(\mathbf{k})}^*=\expval{U_t S_\mu^*U_t^\dagger}_{\mathbf{k}}^*=\expval{T S_\mu T^{-1}}_{\mathbf{k}}^*=-\expval{S_\mu }_{\mathbf{k}}
\end{align}
where we used the TR relations for the unitary matrix in going from the second to the third line. The final equality is obtained by noting that time reversal operating on spin will always bring up a minus sign and since we have an expectation value the complex conjugate acts trivially. Now it is easy to see that the skyrmion curvature is zero since the spin magnitude is unchanged implying:
\begin{align}
    \Omega_{\hat{\mathcal{S}}}(-\mathbf{k})=\hat{\mathcal{S}}(-\mathbf{k}) \cdot\left(\partial_{-k_{x}} \hat{\mathcal{S}}(-\mathbf{k}) \times \partial_{-k_{y}} \hat{\mathcal{S}}(-\mathbf{k})\right)=(-\hat{\mathcal{S}}(\mathbf{k})) \cdot\left(-\partial_{k_{x}} (-\hat{\mathcal{S}}(\mathbf{k}) )\times -\partial_{k_{y}} (-\hat{\mathcal{S}}(-\mathbf{k}))\right)=-\Omega_{\hat{\mathcal{S}}}(\mathbf{k})
\end{align}
Since the skyrmion number is an integral over the whole BZ we obtain the expected result of $\mathcal{Q}=0$ summarizing we have obtained that for a spinful TR system:
\begin{align}
    \expval{S_\mu}_{-\mathbf{k}}=-\expval{S_\mu }_{\mathbf{k}}, \quad \Omega_{\hat{\mathcal{S}}}(-\mathbf{k})=-\Omega_{\hat{\mathcal{S}}}(\mathbf{k}), \quad \mathcal{Q}=0
\end{align}

\section{TR skyrmion invariant}

In this appendix we focus on defining an invariant that characterizes the topology of the TR skyrmion phase given that the total skyrmion number vanishes. To do that we start with the total spin matrix and consider first the system without TR sector coupling. Because of the particle-hole symmetry this operators take the form:

\begin{align}
    S^{I}_\mu=\begin{pmatrix}
    \sigma_\mu  && 0 && 0 && 0 \\
    0 && -\sigma_\mu^* && 0 && 0\\
    0 && 0 && 0 && 0\\
     0 && 0 && 0 && 0
    \end{pmatrix},
    \quad 
       S^{II}_\mu=\begin{pmatrix}
    0  && 0 && 0 && 0 \\
    0 && 0  && 0 && 0\\
    0 && 0 && \sigma_\mu && 0\\
     0 && 0 && 0 && -\sigma_\mu^*
    \end{pmatrix}\label{Spin-operators}
\end{align}

Consider the projector onto the ground state for a given crystal momentum $\textbf{k}$ in the form of:
\begin{align}
    \mathcal{P}_{GS}(\textbf{k})=\sum_n \ketbra{u_n(\textbf{k})}{u_n(\textbf{k})}
\end{align}
where $n=1,\dots n_{occ}$ for $n_occ$ the number of occupied bands. Let he time reversal operator be given by $T=Q K$ where $K$ is complex conjugation and $Q^\dagger=Q^{-1}$,then $T^2=-1$ implies $Q^T=-Q$. It is alsop true that for time reversal systems \cite{Wilsonloop-inversion}:
\begin{align}
    u_n^\alpha(-\textbf{k})=(V_\textbf{k}^{nm})^* Q_{\alpha\beta}(u_m^\beta(\textbf{k}))^*
\end{align}
, where Einstein summation is assumed for $\alpha,\beta$ which run over all components of the vector $\ket{u_n(\textbf{k})}$ and $V_\textbf{k}^{nm}=\mel{u_n(-\textbf{k})}{T}{u_m(\textbf{k})}$ is a sewing matrix between TR partners. It is then easy to prove how the projector unto the Ground state changes under time-reversal:
\begin{align}
    (\mathcal{P}^{\alpha\beta}_{GS}(-\textbf{k}))^*=\sum_n (u_n^\alpha(-\textbf{k})\bar{u}_n^\beta(-\textbf{k}))^*=\sum_n \bar{u}_n^\alpha(-\textbf{k})u_n^\beta(-\textbf{k})=\sum_{n,m} V_\textbf{k}^{nm} Q_{\alpha \gamma}^*u_m^\gamma(\textbf{k})u_n^\beta(-\textbf{k})=\sum_{m} Q^{\beta \alpha'}Q_{\alpha \gamma}^*\bar{u}_m^{\alpha'}(\textbf{k})u_m^\gamma(\textbf{k})
\end{align}
,where we used the fact that $\sum_n V_\textbf{k}^{nm}u_n^\beta(-\textbf{k})=Q^{\beta \alpha} \bar{u}_n^\alpha(\textbf{k})$ so that wit some rearranging we obtain:
\begin{align}
    (\mathcal{P}^{\alpha\beta}_{GS}(-\textbf{k}))^*=\sum_{m} Q_{\alpha \gamma}^*u_m^\gamma(\textbf{k})\bar{u}_m^{\alpha'}(\textbf{k})Q^{\beta \alpha'}=\sum_{m} Q_{\alpha \gamma}^*u_m^\gamma(\textbf{k})\bar{u}_m^{\alpha'}(\textbf{k})(Q^T)^{\alpha'\beta }=\sum_{m} Q^\dagger_{\alpha \gamma}u_m^\gamma(\textbf{k})\bar{u}_m^{\alpha'}(\textbf{k})Q^{\alpha'\beta }
\end{align}
So that we obtain how the projector onto the ground state transforms under time reversal:
\begin{align}
    \mathcal{P}^*_{GS}(-\textbf{k})=Q^\dagger \mathcal{P}_{GS}(\textbf{k})Q
\end{align}
We are interested not on the projector alone but on the pseudo spin texture which may generically be obtained from the expectation value of some spin-like operators $S^{II}_\mu,S^{I}_\mu$ where $\mu$ runs over space indices $\mu=x,y,z$, in our case the pseudo-spin operators may be generically given by $S^{I}_\mu=\tilde{S}_\mu\otimes (1+s_z)/2$
 and $S^{II}_\mu=\tilde{S}_\mu\otimes (1-s_z)/2$ . Here we suppose that the basis is such that there is only matrix elements of $S_\mu^{I,II}$ in one spin sector, where small $s_\mu$ are Pauli matrices representing the spin degrees of freedom. The matrix $\tilde{S}_\mu$ is in principle arbitrary but will turn out to be constrained if TR is satisfied for the textures. We can obtain the pseudo-spin texture by computing the trace:
 \begin{align}
     &\expval{S_\mu^I}_{-\textbf{k}}^*\equiv\text{Tr}(\tilde{S}_\mu\otimes \dfrac{1+s_z}{2}\ \mathcal{P}_{GS}(-\textbf{k}))^*=\text{Tr}(\tilde{S}^*_\mu\otimes \dfrac{1+s_z}{2}\ \mathcal{P}^*_{GS}(-\textbf{k}))=\text{Tr}(\tilde{S}^*_\mu\otimes \dfrac{1+s_z}{2}\ Q^\dagger \mathcal{P}_{GS}(\textbf{k})Q)\\
     &=\text{Tr}(Q\tilde{S}^*_\mu\otimes \dfrac{1+s_z}{2}\ Q^\dagger \mathcal{P}_{GS}(\textbf{k}))=\text{Tr}((-is_y)\tilde{S}^*_\mu\otimes \dfrac{1+s_z}{2}\ (is_y) \mathcal{P}_{GS}(\textbf{k}))=\text{Tr}(\tilde{S}^*_\mu\otimes \dfrac{1-s_z}{2}\ \mathcal{P}_{GS}(\textbf{k}))=\expval{(S_\mu^{II})^*}_{\textbf{k}},
 \end{align}
 where we assumed the unitary part of the TR operator to act only on the spin degree of freedom via $Q=-is_y\otimes \mathbb{I}$ Then considering that the winding number for each pseudo-spin texture can be computed from the effective Hamiltonians:
 \begin{align}
     H_I(\textbf{k})=\expval{S_\mu^I}_{\textbf{k}}\sigma^\mu, \quad H_{II}(\textbf{k})=\expval{S_\mu^{II}}_{\textbf{k}}\sigma^\mu
 \end{align}
 we then have the following relation:
 \begin{align}
     H_I(-\textbf{k})^*=\expval{S_x^I}_{-\textbf{k}}^*\sigma_x^*+\expval{S_y^I}_{-\textbf{k}}^*\sigma_y^*+\expval{S_y^I}_{-\textbf{k}}^*\sigma_z^*=\expval{(S_x^{II})^*}_{\textbf{k}}\sigma_x-\expval{(S_y^{II})^*}_{\textbf{k}}\sigma_y+\expval{(S_z^{II})^*}_{\textbf{k}}\sigma_z
 \end{align}
 We then require the relations $\tilde{S}_x^*=\tilde{S}_x,\ \tilde{S}_z^*=\tilde{S}_z,\ \tilde{S}_y^*=-\tilde{S}_y$ so that TR relates both effective Hamiltonians, in mathematical terms this means:
 \begin{align}
        H_{II}(\textbf{k})=H_I(-\textbf{k})^*
 \end{align}
 We can then construct the decoupled TR model in the pseudo-spin space:
 
\begin{align}
    H_S(\bm{k})\equiv\begin{pmatrix}
    \expval{S^I_\mu}_{\mathbf{k}}\sigma_\mu && 0 \\
    0 && \expval{S^{II}_\mu}_{\mathbf{k}}\sigma_\mu
    \end{pmatrix},\label{Aux-ham}
\end{align}
Interestingly we can calculate the Chern number for each subsector, the result is two skyrmion numbers defined as
\begin{align}
    \mathcal{Q}_I=\frac{1}{4 \pi} \int_{B Z} \hat{\mathcal{S}_I}(\mathbf{k}) \cdot\left(\partial_{k_{x}} \hat{\mathcal{S}_I}(\mathbf{k}) \times \partial_{k_{y}} \hat{\mathcal{S}_I}(\mathbf{k})\right) d^{2} k, \quad 
    \mathcal{Q}_{II}=\frac{1}{4 \pi} \int_{B Z} \hat{\mathcal{S}_{II}}(\mathbf{k}) \cdot\left(\partial_{k_{x}} \hat{\mathcal{S}_{II}}(\mathbf{k}) \times \partial_{k_{y}} \hat{\mathcal{S}_{II}}(\mathbf{k})\right) d^{2} k
\end{align}
,where we have the normalized spin ground state texture for one TR subsector  $I,II$ as $\hat{\mathcal{S}_I}(\mathbf{k})=\mathcal{S}_I(\mathbf{k}) /|\mathcal{S}_{I}(\mathbf{k})|,\hat{\mathcal{S}_{II}}(\mathbf{k})=\mathcal{S}_{II}(\mathbf{k}) /|\mathcal{S}_{II}(\mathbf{k})|$. And $\mathcal{S}_I(\mathbf{k})=\expval{S^I_\mu}_{\mathbf{k}},\mathcal{S}_{II}(\mathbf{k})=\expval{S^{II}_\mu}_{\mathbf{k}}$ since these are the ground state pseudo-spin expectation values. We end up with an uncoupled $\mathbb{Z}_2$ insulator and thus have proved that
\begin{align}
    \mathcal{Q}_I=-\mathcal{Q}_{II},
    \label{Skyrm-I-II-relation}
\end{align}
as long as TR is satisfied for the projector onto the ground state and the pseudo-spin operator satisfies $S^{I}_\mu=\tilde{S}_\mu\otimes (1+s_z)/2$
, $S^{II}_\mu=\tilde{S}_\mu\otimes (1-s_z)/2$ and  $\tilde{S}_x^*=\tilde{S}_x,\ \tilde{S}_z^*=\tilde{S}_z,\ \tilde{S}_y^*=-\tilde{S}_y$. We define in a natural manner the skyrmion parity or TR-skyrmion number as:
\begin{align}
    \nu_{\mathcal{Q}}=(-1)^{\mathcal{Q}_I}=(-1)^{\mathcal{Q}_{II}}=(\mathcal{Q}_I-\mathcal{Q}_{II})/2 \ \  \text{mod}(2)
\end{align}
In addition the previous identity $\expval{S_\mu^I}_{-\textbf{k}}^*=\expval{(S_\mu^{II})^*}_{\textbf{k}}$ also assures us that if somewhere in the BZ the pseudo-spin goes to zero in one spin sector it will also go to zero for the other in the time-reverse momentum. This can be easily verified:
\begin{align}
    |\mathcal{S}_{II}(\mathbf{k})|^2=\expval{S_\mu^{II}}_{\textbf{k}}\expval{S_\mu^{II}}_{\textbf{k}}=\expval{(S_\mu^{II})^*}_{\textbf{k}}\expval{(S_\mu^{II})^*}_{\textbf{k}}=\expval{S_\mu^{I}}_{-\textbf{k}}^*\expval{S_\mu^{I}}_{-\textbf{k}}^*=\expval{S_\mu^{I}}_{-\textbf{k}}\expval{S_\mu^{I}}_{-\textbf{k}}= |\mathcal{S}_{I}(-\mathbf{k})|^2
\end{align}
,where we used the previous derived identity as well as the properties of the operators that need to be satisfied in order for TR to relate them, we obtain then:
\begin{align}
    |\mathcal{S}_{I}(\mathbf{k})|^2= |\mathcal{S}_{II}(-\mathbf{k})|^2,\quad \text{min}_{\text{BZ}}{\mathcal{S}_I(\mathbf{k})}=\text{min}_{\text{BZ}}{\mathcal{S}_{II}(\mathbf{k})}
\end{align}
, in other words the topological information of both textures can be inferred from only one pseudo-spin texture. The last equality implies that a phase transition occurring when one pseudo-spin magnitude goes to zero destabilizes the other pseudo-spin sector such that both skyrmion numbers jump simultaneously.

\section{Observable-enriched partial trace}

In this section we derive the relationship between the skyrmion texture and a partial trace of the ground state projector. We begin with the previous expression in terms of the total trace:
\begin{align}
    &\expval{S_\mu^I}_{\textbf{k}}\equiv\text{Tr}\left\{\mathcal{P}_{GS}(\textbf{k}) \tilde{S}_\mu\otimes \dfrac{1+s_z}{2}\ \right\}=\text{Tr}\left\{U^\dagger U \mathcal{P}_{GS}(\textbf{k}) U^\dagger U \tilde{S}_\mu\otimes \dfrac{1+s_z}{2}\ \right\}\\
    &=\text{Tr}\left\{U \mathcal{P}_{GS}(\textbf{k}) U^\dagger (U_1\otimes \mathbb{I}) \tilde{S}_\mu\otimes \dfrac{1+s_z}{2}\ (U_1\otimes \mathbb{I} )^\dagger  \right\}= \text{Tr}\left\{U \mathcal{P}_{GS}(\textbf{k}) U^\dagger (U_1\tilde{S}_\mu U_1^\dagger \otimes \mathbb{I})(\mathbb{I}\otimes  \dfrac{1+s_z}{2})\  \right\}
\end{align}
,where we introduced the unitary operator $U=U_1\otimes \mathbb{I}$ where the identity is in spin space and we select $U_1=\begin{pmatrix}1 && 0 \\
0 && \sigma_y\end{pmatrix}$ which rotates the pseudo-spin operators so that now:
\begin{align}
    \expval{S_\mu^I}_{\textbf{k}}=\text{Tr}\left\{U \mathcal{P}_{GS}(\textbf{k}) U^\dagger (\sigma_\mu \otimes \mathbb{I}) P_\uparrow\  \right\}=\text{Tr}\left\{ P_\uparrow U \mathcal{P}_{GS}(\textbf{k}) U^\dagger (\sigma_\mu \otimes \mathbb{I}) \  \right\}=\text{Tr}
    \left\{\rho_\uparrow \sigma_\mu \  \right\}
\end{align}
, where $P_\uparrow$ projects to the up spin sector, we used cyclicity of the trace and defined the reduced density matrix as the matrix such that the last equality is always fulfilled. The last trace is a trace only in the spin space no longer containing other degrees of freedom,except $\sigma$. The operators $\rho_\uparrow$  and $\rho_\downarrow$, for which we can apply the same reasoning are defined as:
\begin{align}
    &\rho_\uparrow (\textbf{k})= \text{Tr}_{\bar{\sigma}}\left\{ P_\uparrow U \mathcal{P}_{GS}(\textbf{k}) U^\dagger \right\}\\
    &\rho_\downarrow(\textbf{k}) = \text{Tr}_{\bar{\sigma}}\left\{ P_\downarrow U \mathcal{P}_{GS}(\textbf{k}) U^\dagger \right\}
\end{align}

This textures are linked then to the pseudo-spin expectation value in a simple way. By using the fact that $\rho_\uparrow (\textbf{k})=d_0(\textbf{k})\sigma_0+d_\nu(\textbf{k})\sigma_\nu$ (Einstein summation implied) and $\rho_\downarrow(\textbf{k})$ are 2 by 2 hermitian matrices generated by Pauli matrices one finds:
\begin{align}
    \expval{S_\mu^I}_{\textbf{k}}=\text{Tr}\left\{d_0(\textbf{k})\sigma_0 \sigma_\mu+d_\nu(\textbf{k})\sigma_\nu \sigma_\mu\right\}=2 d_\mu(\textbf{k})
\end{align}
It is clear then that the pseudo-spin topology is equivalent to the topology of the reduced density matrix for the pseudo-spin subsector defined above, with the final relation:
\begin{align}
    \rho_\uparrow (\textbf{k})=d^\uparrow _0(\textbf{k})\sigma_0 +\frac{1}{2}\expval{S_\mu^I}_{\textbf{k}}\sigma_\mu\\
     \rho_\downarrow (\textbf{k})=d^\downarrow _0(\textbf{k})\sigma_0 +\frac{1}{2}\expval{S_\mu^{II}}_{\textbf{k}}\sigma_\mu
\end{align}
This relationship can be tested numerically and we found an excellent agreement between both textures, an example texture is shown in Fig. \label{rho_up}.

\begin{figure}[ht]
\includegraphics[width=\columnwidth,height=4.8cm]{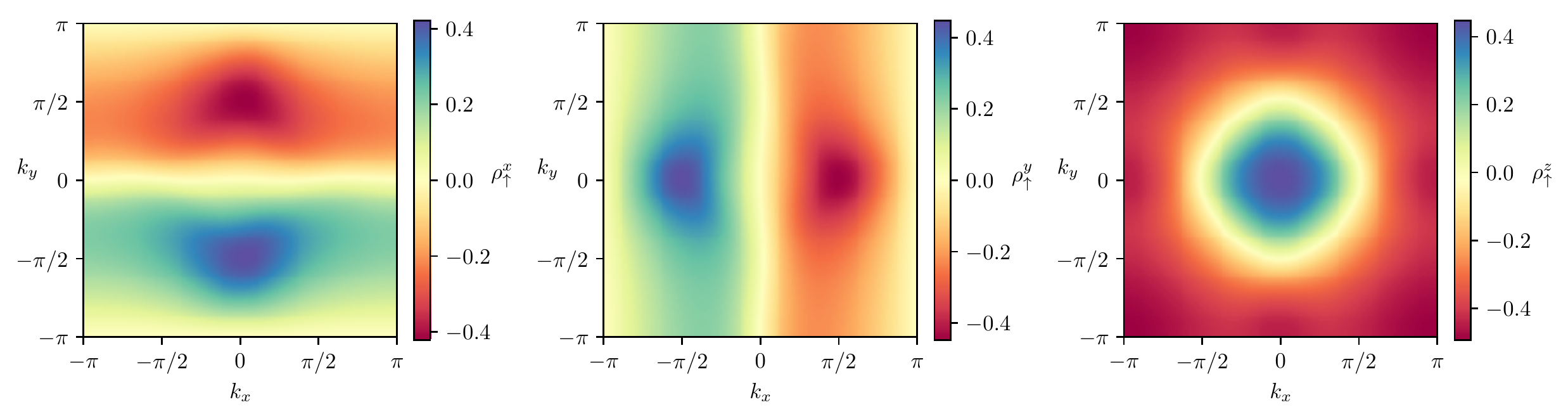}
  \vspace{-0.3cm}
  \caption{ $\rho_\uparrow (\textbf{k})$ reduced density matrix components as a function of crystal momentum for the model of \eqref{Skyrm-mod1} with $M=-1$, $\Delta_0=0.5$,$\lambda=0.3$, $c=0.5$. }
  \label{rho_up}
\end{figure}

Additionally the link discussed in the main text between the reduced entanglement spectrum or reduced 2-point correlator is clear from the following derivation. We start with the known non-interacting result for the 2-point correlator which states that the correlation matrix is given by \cite{Bernevig-entanglement}:
\begin{align}
G_{n m}^{\alpha \beta}\left(k_x\right) =\left\langle\Psi_{G S}\left|c_{k_x n \alpha}^{\dagger} c_{k_x m \beta}\right| \Psi_{G S}\right\rangle =\sum_{k_y}\sum_{\tilde{n}\in \text{occ}}\mathcal{P}^{\alpha \beta *}_{\tilde{n}}(\textbf{k})e^{i(n-m)k_y},
\end{align}
where $\sum_{\tilde{n}\in \text{occ}}\mathcal{P}^{\alpha \beta *}_{\tilde{n}}(\textbf{k})$ gives the matrix elements for the ground state projector $\mathcal{P}_{\text{GS}}(\textbf{k})$ of the fully periodic boundary Hamiltonian and we have expressed the real space  $y$ correlation matrix as $G_{n m}^{\alpha \beta}\left(k_x\right)$. Here $c_{k_x n \alpha}^{\dagger}$ creates an electron with momentum $k_x$ in the $n$ lattice site and in state $\alpha$. Where $\alpha$ is the local Hilbert space in each site which for us consists of particle-hole,spin and pseudo-spin. Next let us multiply by the unitary considered above and project to the up spin sector so that one obtains the fourier transform of  $ P_\uparrow U \mathcal{P}_{GS}(\textbf{k}) U^\dagger $. Then proceeding to take the trace over all other degrees of freedom except pseudo-spin $\sigma$ we have:

\begin{align}
\left[\tilde{G}_\uparrow\right]_{n m}^{\sigma \sigma' }\left(k_x\right) = \left[ \text{Tr}_{\bar{\sigma}}\{P_\uparrow U G_{n m}\left(k_x\right)U^\dagger\}\right]^{\sigma \sigma' } =\sum_{k_y} \left[ \rho_{\uparrow}(\textbf{k})\right]^{\sigma \sigma' }e^{i(n-m)k_y},
\end{align}

Furthermore let us trace over $N_y-L$ lattice sites, subsystem $B$, so that we have re reduced correlation matrix which will be a Toepplitz matrix of dimension $d_{\text{loc}}*L$, subsytem $A$, which we denote $\tilde{G}_\uparrow^A(k_x)$. Now because there is no periodicity in $L<N_y$ we have an open boundary condition matrix which we now wish to diagonalize. Since the elements of this matrix are all given by the real space transform of $\rho_{\uparrow}(\textbf{k})^{\sigma \sigma' }$ with no coupling between site $L$ and site $1$, for $L$ large enough so that no nearest neighbour hopping exists, we have just the open boundary of the matrix $d^\uparrow _0(\textbf{k})\sigma_0 +\frac{1}{2}\expval{S_\mu^I}_{\textbf{k}}\sigma_\mu$. Since the Chern number of $\rho_{\uparrow}(\textbf{k})$ is just the Skyrmion invariant and applying the ten-fold way bulk-boundary correspondence for $\rho_{\uparrow}(\textbf{k})$ we conclude that the spectrum of $\tilde{G}_\uparrow^A(k_x)$ will contain $\mathcal{Q}_{\uparrow}$ chiral gapless states which cannot be gapped out by any perturbation. The same reasoning leads to $\mathcal{Q}_{\downarrow}$ chiral gapless states for the down projected reduced correlation matrix.\\

To visualize the $\mathbb{Z}_2$ nature of the invariant we can repeat the same argument but now not taking the projector and tracing out only the particle-hole d.o.f. we obtain the same relation but with $\text{Tr}_{\tau}\{U\mathcal{P}_{GS}(\textbf{k})U^\dagger\}$ which is now a 4 by 4 matrix in spin and pseudo-spin d.o.f as such the upper 2 by 2 component is just the matrix $\bra{\uparrow,\sigma}\text{Tr}_{\tau}\{U\mathcal{P}_{GS}(\textbf{k})U^\dagger\}\ket{\uparrow,\sigma}$  and the down-spin matrices. These components are related by TR and tied to the Skyrmion invariant since the following relation holds:
\begin{align}
    \bra{\uparrow,\sigma}\text{Tr}_{\tau}\{U\mathcal{P}_{GS}(\textbf{k})U^\dagger\}\ket{\uparrow,\sigma'}&=\bra{\sigma}\text{Tr}_{s}P_{\uparrow}\text{Tr}_{\tau}\{U\mathcal{P}_{GS}(\textbf{k})U^\dagger\}\ket{\sigma'}=\bra{\sigma} \text{Tr}_{s}\text{Tr}_{\tau}\{P_{\uparrow}U\mathcal{P}_{GS}(\textbf{k})U^\dagger\}\ket{\sigma'}\notag \\
    &=\bra{\sigma} \text{Tr}_{\bar{\sigma}}\{P_{\uparrow}U\mathcal{P}_{GS}(\textbf{k})U^\dagger \}\ket{\sigma'}=\rho^{\uparrow}_{\sigma \sigma'}(\textbf{k})
\end{align}
similar relation holds for the down component and thus both matrices are related by TR symmetry as proven above for the reduced density matrices of the bulk. Moreover, the invariant for each sector $\mathcal{Q}_{\uparrow}=-\mathcal{Q}_{\downarrow}$ is then proven to be linked to the topology of $\text{Tr}_{\tau}\{U\mathcal{P}_{GS}(\textbf{k})U^\dagger\}$ since the $\mathbb{Z}_2$ invariant in this case is just $\nu_Q=(-1)^{\mathcal{Q}_{\uparrow}}$. Since we still have a non-diagonal SOC matrix element $\bra{\uparrow,\sigma}\text{Tr}_{\tau}\{U\mathcal{P}_{GS}(\textbf{k})U^\dagger\}\ket{\downarrow,\sigma'}$ one would need to calculate the Kane-Mele index for the 4 by 4 matrix $\text{Tr}_{\tau}\{U\mathcal{P}_{GS}(\textbf{k})U^\dagger\}$ if only time-reversal symmetry is conserved in the subsystem. In this sense applying the reasoning of the previous chapter for the correlation matrix we can see that the helical edge states of opening boundary conditions for $\text{Tr}_{\tau}\{U\mathcal{P}_{GS}(\textbf{k})U^\dagger\}$ results in a gapless reduced correlation spectrum from $\left[\tilde{G}_{\nu}(k_x)\right]^{\sigma s ; \sigma' s' }_{nm}=  \text{Tr}_{\tau}\{U G_{n m}\left(k_x\right)U^\dagger\}^{\sigma s ; \sigma' s' } $ protected solely by time-reversal symmetry.

The robustness of the reduced entanglement spectrum or equivalently of the reduced correlator spectrum which is shown in the main text against all TR and PH symmetry allowed perturbations indicates that the subsystem only needs the previous symmetries in the full system to maintain TR symmetry. This is clear if the matrix $U$ is the identity since then $\Tr_\tau{\rho(\bm{k})}$ will also have TR symmetry just by using equation (S15) and that the trace does not involve the $W$ matrix so it can get out of the partial trace. The case of an observable which includes a different unitary matrix needs to be treated more delicate since the unitary matrix will mix and entangle degrees of freedom involved in the partial trace. Specializing to our case with $U=(1-\sigma_y)\tau_z/2+(1+\sigma_y)\tau_0/2$ we see that the trace now becomes: 
\begin{align}
    \Tr_\tau{U\rho(\bm{k})U^\dagger}= \sum_{\tau,\sigma,s,s',\sigma'} \rho^{\sigma' s' \tau}_{\sigma s \tau}(\tau \delta_{\sigma',-1}+\delta_{\sigma',+1})(\tau \delta_{\sigma,-1}+\delta_{\sigma,+1})\ketbra{\sigma s}{\sigma ' s'}=\sum_{\sigma,\sigma',\tau} A^\tau_{\sigma,\sigma',\tau}\ketbra{\sigma}{\sigma'}\begin{pmatrix}
    \rho^{\sigma' \uparrow \tau}_{\sigma \uparrow \tau}\ketbra{\uparrow}{\uparrow} && \rho^{\sigma' \uparrow \tau}_{\sigma \downarrow \tau}\ketbra{\uparrow}{\downarrow} \\
    \rho^{\sigma' \downarrow \tau}_{\sigma \uparrow \tau}\ketbra{\downarrow}{\uparrow} && \rho^{\sigma' \downarrow \tau}_{\sigma \downarrow \tau}\ketbra{\downarrow}{\downarrow}
\end{pmatrix}
\end{align}
,where $\rho^{\sigma' s' \tau}_{\sigma s \tau}$ are the matrix elements  of the traced out ground state projector with momentum dependence being implicit, and $A^\tau_{\sigma,\sigma'}=(\tau \delta_{\sigma',-1}+\delta_{\sigma',+1})(\tau \delta_{\sigma,-1}+\delta_{\sigma,+1})$ come from the form of the unitary matrix $U$ considered before in the basis of $\sigma_y$ eigenvectors. Now requiring time reversal symmetry for this reduced ground state projector implies a fix form and relation between the matrix elements for different momentum values. The diagonal terms for example need to fulfill the usual relationship of a TR hamiltonian meaning:
\begin{align}
    \sum_\tau A^\tau_{\sigma,\sigma'}\rho^{\sigma' \uparrow \tau}_{\sigma \uparrow \tau}(\bm{k})= \sum_\tau  A^\tau_{\sigma,\sigma'}\rho^{\sigma' \downarrow \tau}_{\sigma \downarrow \tau}(-\bm{k})^*
\end{align}
this is clearly fullfilled by the TR invariance of the full ground state projector if $\sigma=\sigma'$ since then the $A$ tensor is the identity. For the antidiagonal case not much can be said unless there is a relationship involving $\tau \rho(\bm{k})$ which indicates that a particle hole symmetry of the full system is needed since PH always results in an additional negative sign and a reversal of spin. Finally the same reasoning works for the antidiagonal components since TR symmetry implies:
\begin{align}
    \sum_\tau A^\tau_{\sigma,\sigma'}\rho^{\sigma' \uparrow \tau}_{\sigma \downarrow \tau}(\bm{k})=\sum_\tau  -A^\tau_{\sigma,\sigma'}\rho^{\sigma' \downarrow \tau}_{\sigma \uparrow \tau}(-\bm{k})^*
\end{align}
,this can be fulfilled trivially for the $\sigma=\sigma'$ components by using the TR symmetry of the full system and for the off diagonal in $\sigma$ the relation will be fulfilled with the additional PH symmetry of the full system which changes the $\tau$ sign coming from the $A$ tensor.

\section{Bulk-boundary correspondence and symmetry protection}

As discussed in the main text, the edge states encountered for a non-trivial skyrmion parity remain gapless even in the presence of non-negligible spin-orbit coupling and a bulk perturbation breaking crystalline point group symmetries. Here, we investigate in detail the symmetry protection of the edge states and the role of spin-orbit coupling. As a starting point we impose that the system is invariant under $\mathcal{T}=-is_y K$ and $\mathcal{C}=s_z\tau_x \sigma_z K$. The allowed momentum even spin-orbit terms that couple both up and down spin are then reduced to just six terms:

\begin{align}
    V_1=s_x\tau_x \sigma_y ,\quad V_2= s_x\tau_y \sigma_z , \quad V_3= s_y \tau_z \sigma_y, \notag \\
    B_1=s_x \tau_0 \sigma_y,\quad B_2=s_x\tau_y \sigma_0, \quad B_3=s_y \tau_y \sigma_x 
\end{align}

Out of these six terms, it is numerically verified that $B_1,B_2,B_3$ do not gap out the edge states when open boundary conditions are considered, even in the presence of $V_{\text{bulk}}(\textbf{k})$ which breaks all crystalline point group symmetries. The remaining terms---although respecting symmetries of the full Hamiltonian---do not necessarily respect symmetries enforced on subsets of the degrees of freedom, corresponding to sub-matrices of the full Bloch Hamiltonian representation. Additionally although TR is maintained as long as TR and PH is present in the full system this doesn't mean that the edge states of the full system have a protected connection to the topology of the subset of degrees of freedom we consider, since Energy and spin are generally decoupled. It is in this context that such subset symmetry-protection which has not been considered previously, to our knowledge, is relevant to the TRI topological skyrmion phase. The skyrmion invariant is essentially considering topology of the $\sigma$ degree of freedom, so enforcing symmetries on this subset does result in $V_i$ terms being ruled out and therefore protects the gapless helical edge modes, which survive for non-trivial $\nu_{\mathcal{Q}}$ and trivial $\nu$. \\

To simplify discussion of these symmetries enforced on subsets of the degrees of freedom, we reformulate generic SOC terms as:

\begin{align}
    V=\begin{pmatrix}
    0 && 0 && V_{pp} && V_{ph} \\
     0 && 0 && V_{hp} && V_{hh} \\
     V^\dagger_{pp} && V^\dagger_{hp} && 0 && 0 \\
     V^\dagger_{ph} && V^\dagger_{hh} && 0 && 0 \\
    \end{pmatrix},
\end{align}

where each element in $V$ has $2\times2$ matrix representation. First, we define the $\mathcal{C}'$-symmetry enforced only on the $\tau$ and $\sigma$ dofs. Defining the upper right block of $V$ with $4 \times 4$ matrix representation as:

\begin{align}
    V_{\text{SOC}}=\begin{pmatrix}
    V_{pp} && V_{ph} \\
     V_{hp} && V_{hh} \\
    \end{pmatrix},
\end{align}
its invariance under the $\mathcal{C}'$ transformation is specified by requiring:
\begin{align}
    J_{\tau \sigma}V_\text{SOC}^*J_{\tau \sigma}^{-1}=-V_\text{SOC}
\end{align}

where $J_{\tau \sigma}= -i \tau_y \sigma_0$. By plugging in the six allowed SOC terms we see that $V_1$ and $V_2$ break this symmetry, while $B_1$, $B_2$, and $B_3$ each respect it. $V_3$ also respects this symmetry, indicating a second, previously-unidentified symmetry protecting the more robust helical edge modes is broken by $V_3$. We consider the symmetry-protection required in the full system to ensure time-reversal symmetry in the spin subsystem generated by performing a observable-enriched partial trace over the $\tau$ degree of freedom. We therefore extract particle coupling only, which will have elements of $H_0$ the uncoupled Hamiltonian in the diagonal and on the off-diagonal a matrix:

\begin{align}
    V_p=\begin{pmatrix}
    0 && V_{pp}  \\ \\
     V^\dagger_{pp} && 0 \\
    \end{pmatrix}
\end{align}

We define the constraint for a chiral symmetry of $V_p$, denoted by the operation $\left(\mathcal{C}\mathcal{T} \right)_{p}$ as:

\begin{align}
   \Omega_{p}V_{p}\Omega_{p}^{-1}=-V_{p},
\end{align}

where $\Omega_{p} = s_x \sigma_z$, which comes from restricting the original particle-hole symmetry $\mathcal{C} $ times the time-reversal operator $\mathcal{T}$ to the particle subspace. By explicit computation it is easy to show that $V_3$ breaks this $\left(\mathcal{C}\mathcal{T} \right)_{p}$ symmetry, while all other terms $V_1,V_2,B_1,B_2,B_3$ do not. Consistently the diagonal elements from the full Hamiltonian restricted to this subspace also do not break the combined $\left(\mathcal{C}\mathcal{T} \right)_{p}$. We may thus understand the additional robustness of the helical edge modes as deriving in part from the realization of a QSHI phase of the spin subsystem.

\begin{figure}[ht!]
\subfigure[]{\includegraphics[width=0.47\columnwidth,height=5.5cm]{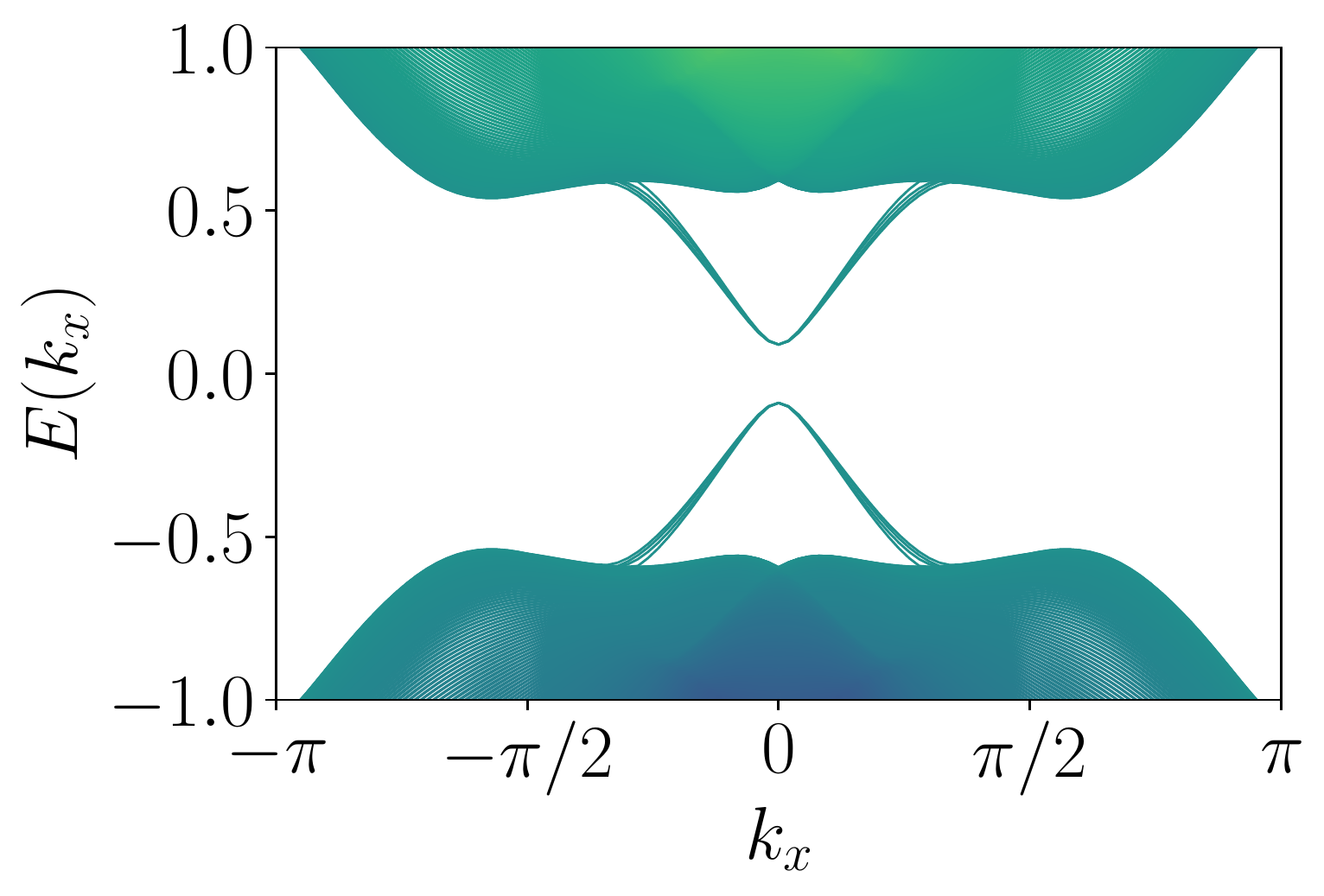}}
\subfigure[]{\includegraphics[width=0.47\columnwidth,height=5.5cm]{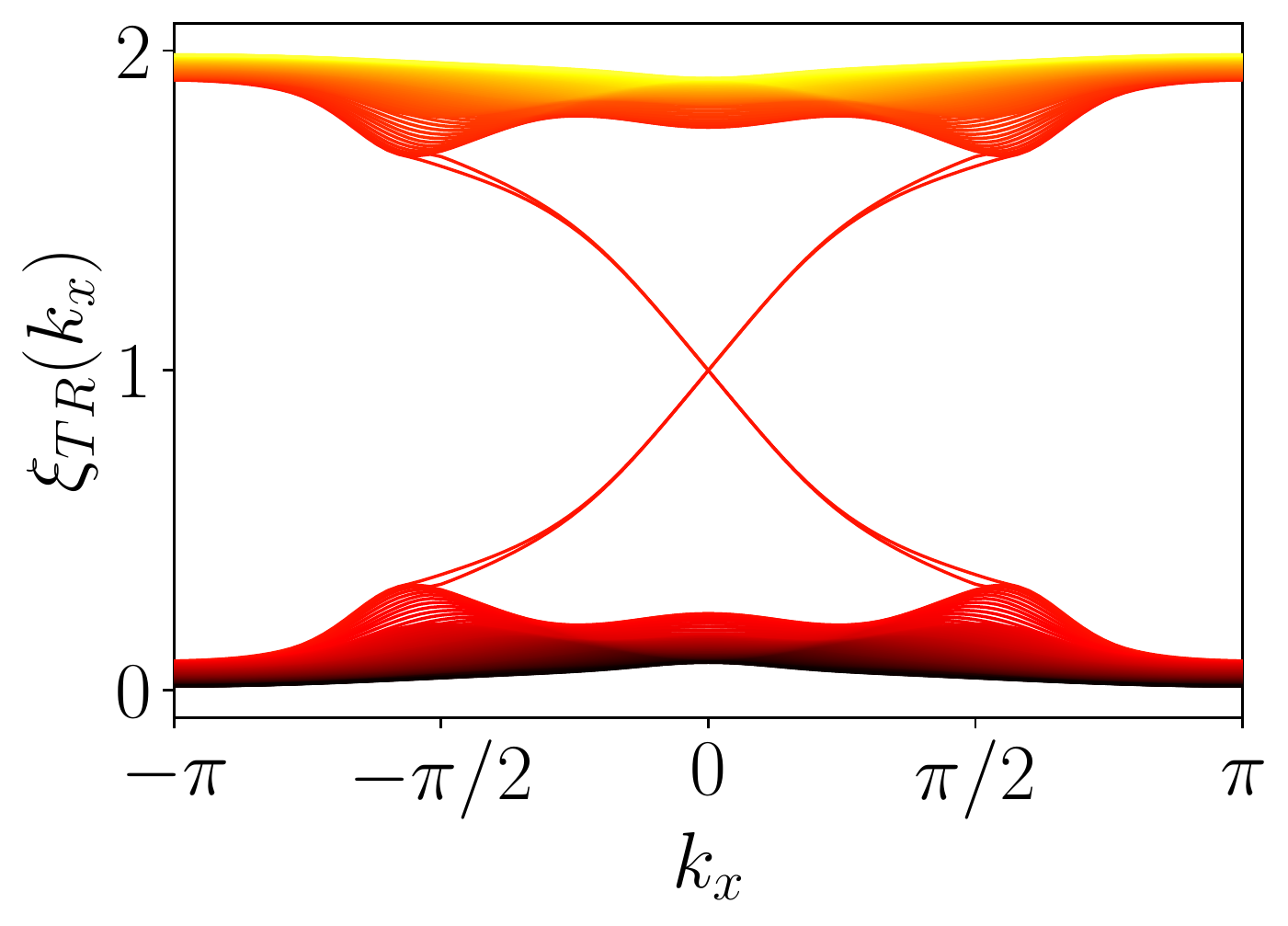}}\\
\subfigure[]{\includegraphics[width=0.32\columnwidth,height=4cm]{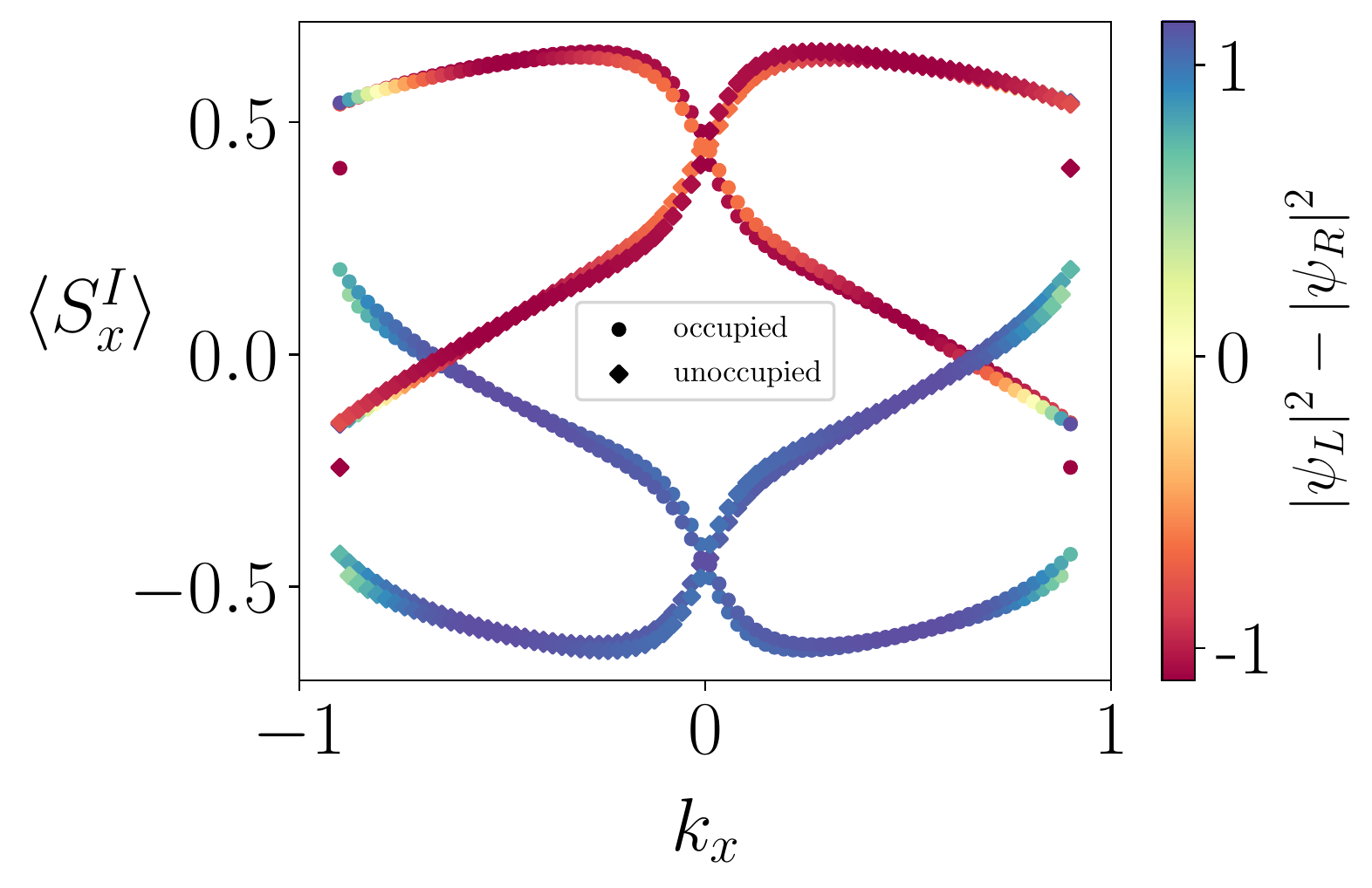}}
\subfigure[]{\includegraphics[width=0.32\columnwidth,height=4cm]{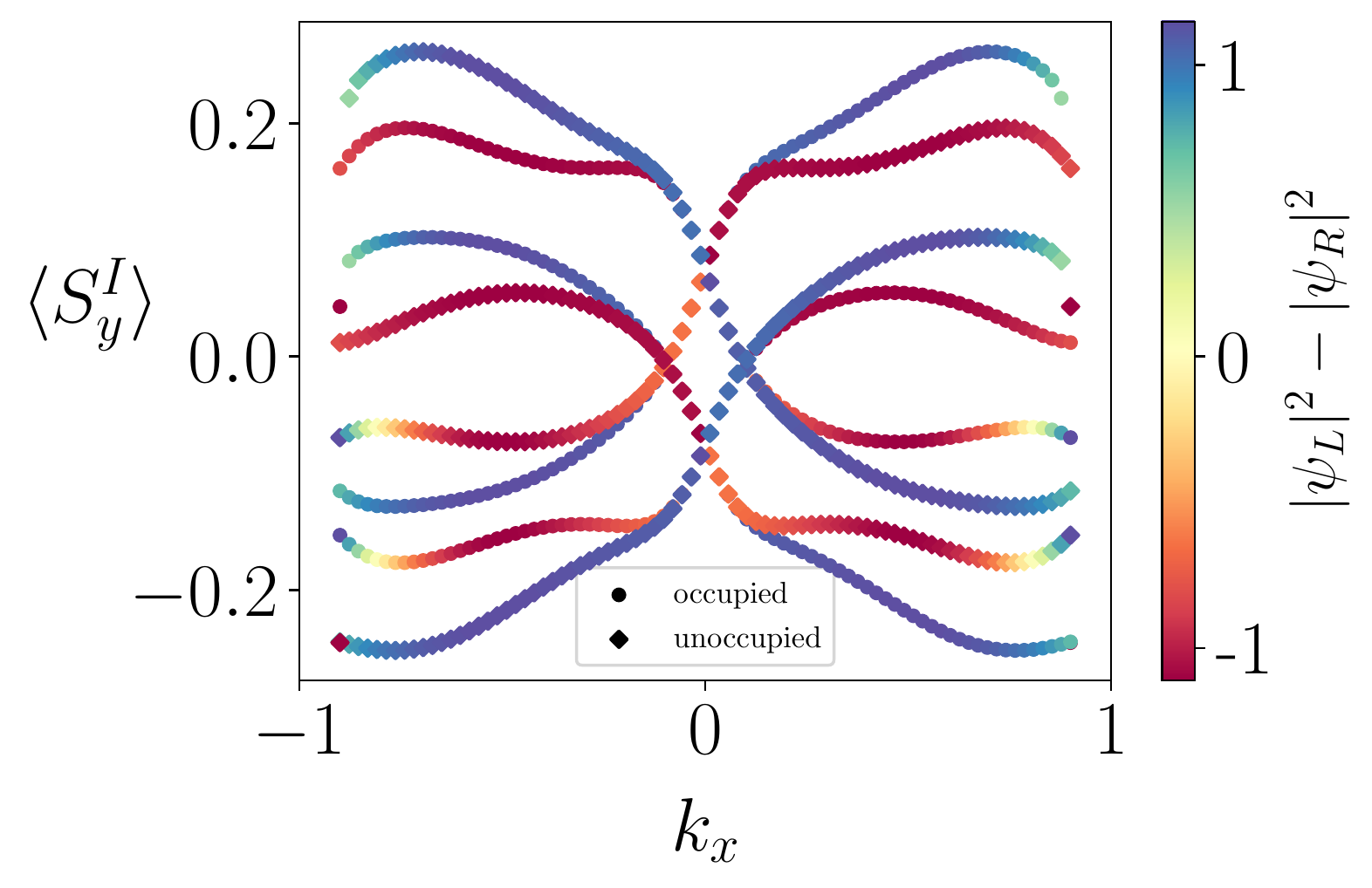}} 
 \subfigure[]{\includegraphics[width=0.32\columnwidth,height=4cm]{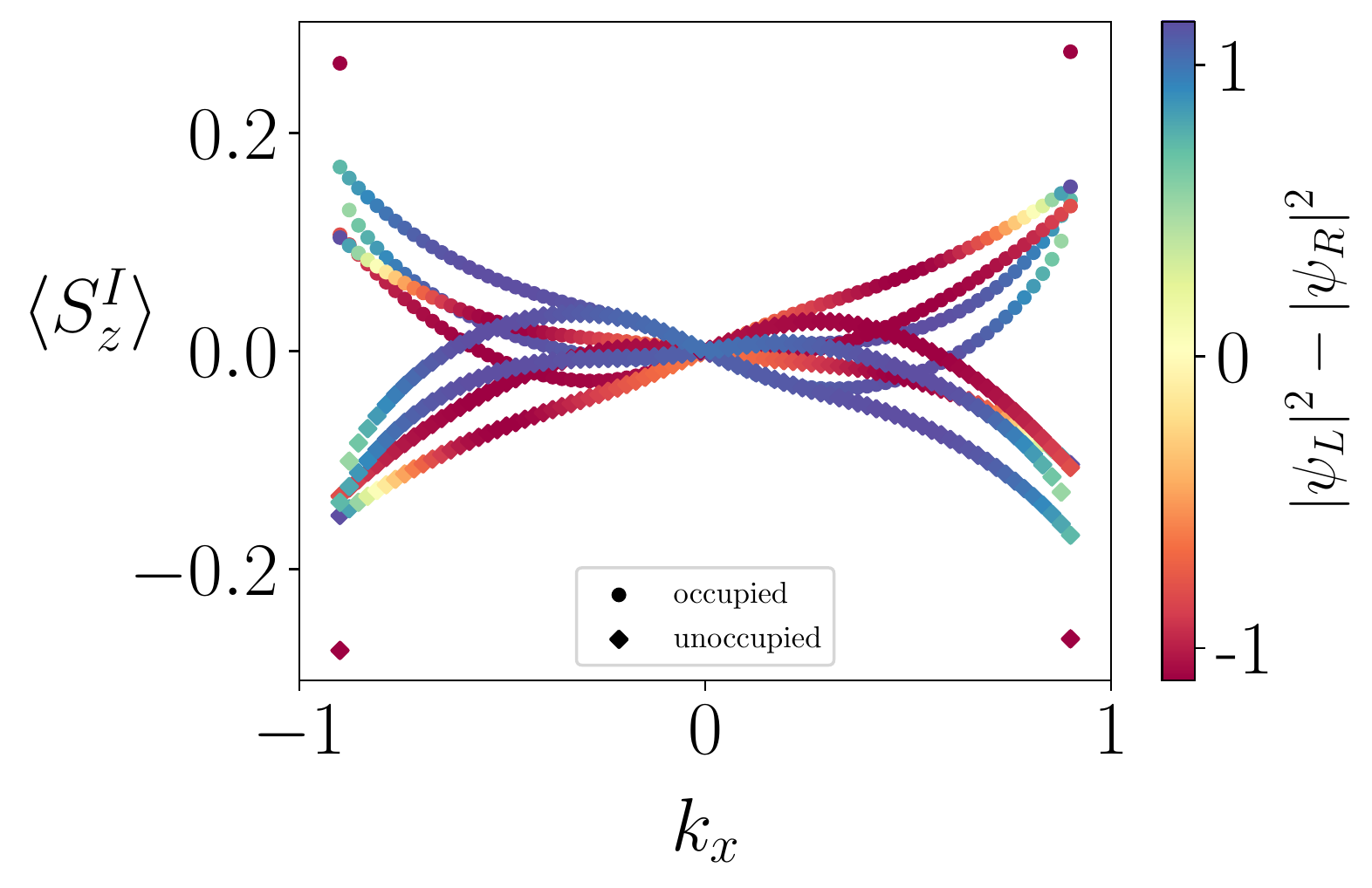}}
 
 \hspace{0.4cm}
     
  \vspace{-0.3cm}
  \caption{a) Energy spectrum for the model of Eq. \eqref{Skyrm-mod1} with OBC in $y$ and periodic in $x$ with parameters $M=-1$, $\Delta_0 = 0.5$,$c=0.5$ length $N_y=200$ and $\lambda=0.3$ with the edge perturbation discussed in the text $V_3$ of strength $0.2$. b) 2-point reduced correlation spectrum for the Hamiltonian in equation \ref{Skyrm-mod1} with parameters $M = -1$,$\Delta_0 = 0.5$, $\lambda = 0.3$ , $c = 0.5$ and perturbation $V_3$ of strength $0.2$ turned on, for a cut along the $y$ axis and traced over $L=200$ lattice sites and particle-hole d.o.f. c) $x$ component of the pseudo-spin expectation value for the edge states of the Hamiltonian in equation \ref{Skyrm-mod1} with parameters $M = -1$,$\Delta_0 = 0.5$, $\lambda = 0.3$ , $c = 0.5$ and open boundary conditions on $y$. The perturbation $V_3$ was added with amplitude $0.2$. The color is proportional to the localization on either edge computed by taking the probability density at each edge and subtracting them. d) and e) show the $y,z$ components of the pseudo-spin expectation value for the edge states respectively. }
  \label{Tp-plots}
\end{figure}

\end{document}